\UseRawInputEncoding
\documentclass[fleqn,usenatbib]{mnras}
\pdfoutput=1
\usepackage{hyperref}
\usepackage{newtxtext,newtxmath}
\usepackage[dvipsnames]{xcolor}

\usepackage[T1]{fontenc}

\DeclareRobustCommand{\VAN}[3]{#2}
\let\VANthebibliography\thebibliography
\def\thebibliography{\DeclareRobustCommand{\VAN}[3]{##3}\VANthebibliography}

\usepackage{graphicx}	
\usepackage{amsmath}	
\usepackage{longtable}
\usepackage{orcidlink}
\usepackage{pdfpages} 
\usepackage{pgffor} 
\usepackage{bibunits}

\usepackage{pdfpages} 
\usepackage{pgffor} 

\makeatletter
\AtBeginDocument{\let\LS@rot\@undefined}
\makeatother

\def\supplementfilenameA{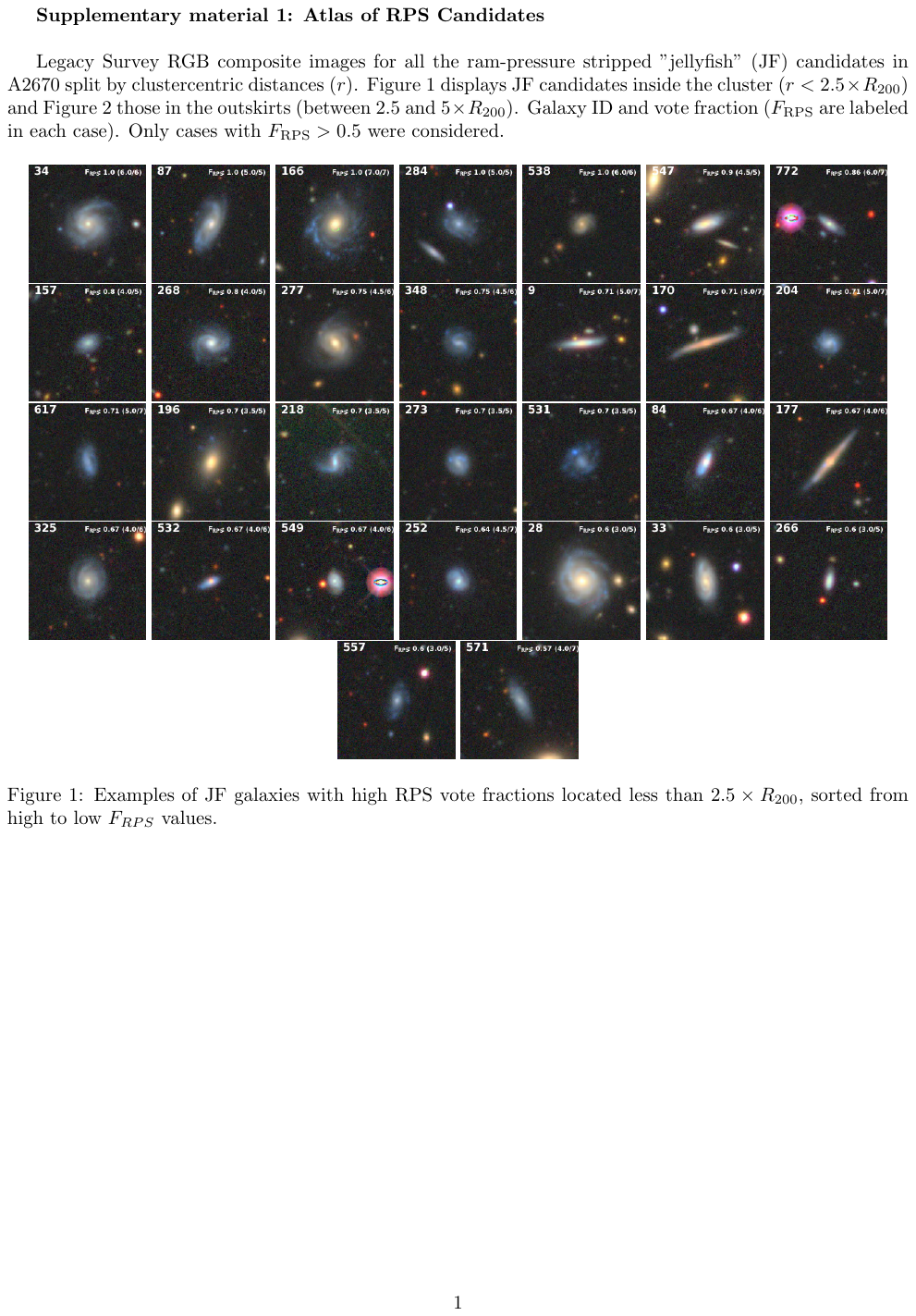}
\def\supplementfilenameB{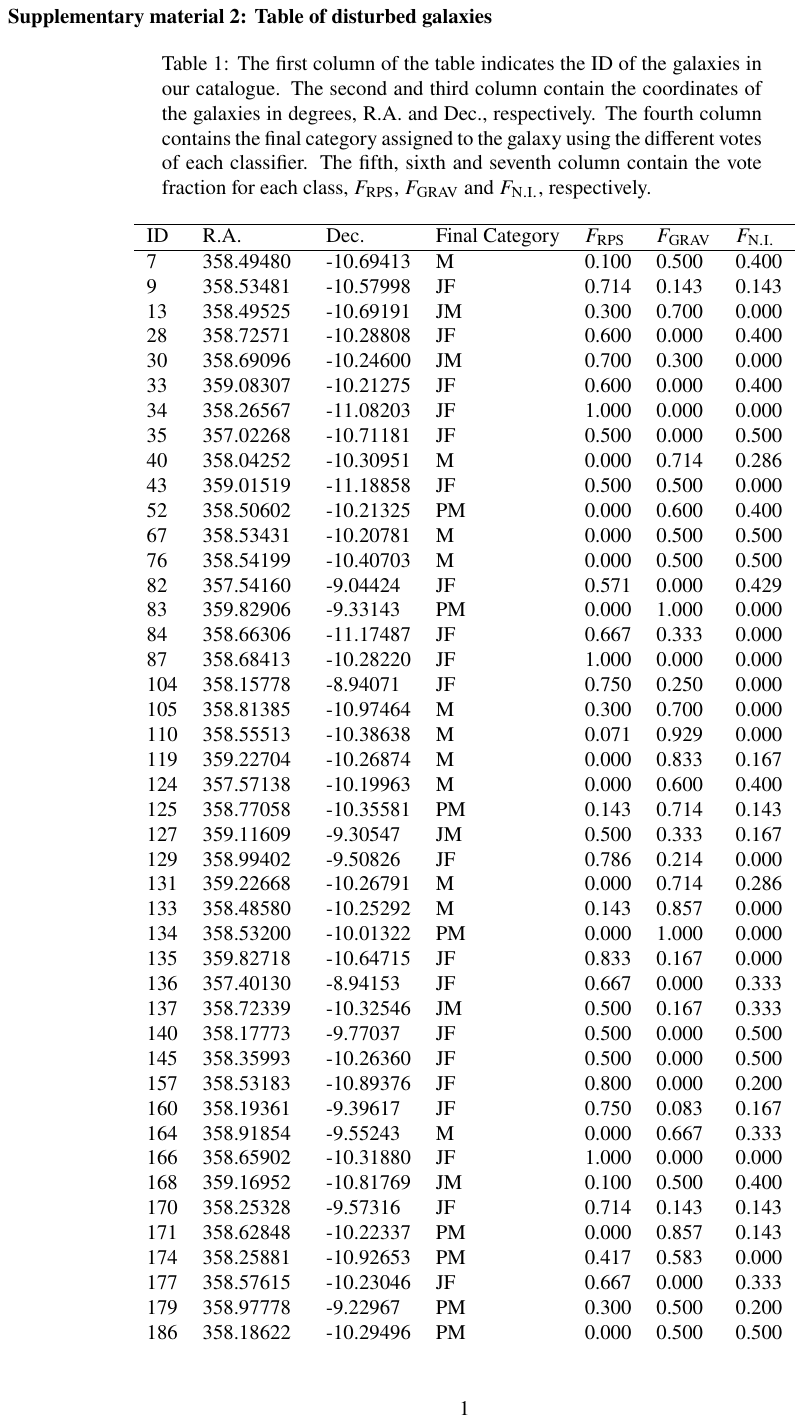}

\pdfximage{\supplementfilenameA}
\def\numbersupplementpagesA{\the\pdflastximagepages}

\pdfximage{\supplementfilenameB}
\def\numbersupplementpagesB{\the\pdflastximagepages}

\newif\ifarXiv
\arXivtrue 

\title[Pre- and post-processing of cluster galaxies]{Pre- and post-processing of cluster galaxies out to $5 \times R_{200}$:\\ The extreme case of A2670}

\author[Piraino-Cerda et al.]{
Franco Piraino-Cerda,$^{1,2}\orcidlink{0009-0008-0197-3337}$
Yara L. Jaff\'e,$^{1}\orcidlink{0000-0003-2150-1130}$\thanks{E-mail: yara.jaffe@usm.cl}
Ana C. Louren\c{c}o,$^{2,3}\orcidlink{0000-0002-4393-7798}$
Jacob P. Crossett,$^{1,2}\orcidlink{0000-0002-9810-1664}$
Vicente Salinas,$^{2}\orcidlink{0000-0002-6858-4976}$
\newauthor
Duho Kim,$^{4,5}\orcidlink{0000-0001-5120-0158}$
Yun-Kyeong Sheen,$^{5}\orcidlink{0000-0002-3211-9431}$
Kshitija Kelkar,$^{1,2}\orcidlink{0000-0002-8130-3593}$
Diego Pallero,$^{1}\orcidlink{0000-0002-1577-7475}$ and 
Hector Bravo-Alfaro$^{6}$
\\
$^{1}${Departamento de F\'isica, Universidad T\'ecnica Federico Santa Mar\'ia, Avda. Espa\~na 1680, Valpara\'iso, Chile}\\
$^{2}${Instituto de F\'isica y Astronom\'ia, Universidad de Valpara\'iso, Avda. Gran Breta\~na 1111, Valpara\'iso, Chile}\\
$^3${European Southern Observatory, Alonso de Cordova 3107, Santiago, Chile}\\
$^{4}${Department of Astronomy and Space Science, Chungnam National University, 99 Daehak-ro, Daejeon 305-764, Republic of Korea}\\
$^5${Korea Astronomy and Space Science Institute, Daedeokdae-ro 776, Yuseoung-gu Daejeon 34055, Republic of Korea}\\
$^6${Departamento de Astronom\'ia, Universidad de Guanajuato, Guanajuato 36000, Mexico}
}

\date{Accepted XXX. Received YYY; in original form ZZZ}

\pubyear{2023}

\begin{document}
\label{firstpage}
\pagerange{\pageref{firstpage}--\pageref{lastpage}}
\maketitle

\begin{abstract}
We study galaxy interactions in the large scale environment around A2670, a massive ($M_{200}$ = $8.5 \pm 1.2~\times 10^{14} \mathrm{M_{\odot}}$) and interacting galaxy cluster at z = 0.0763. 
We first characterize the environment of the cluster out to 5$\times R_{200}$ 
and 
find a wealth of substructures, including the main cluster core, a large infalling group, and several other substructures. To study the impact of these substructures (pre-processing) and their accretion into the main cluster (post-processing) on the member galaxies, we visually examined 
optical images to look for signatures indicative of gravitational or hydrodynamical interactions. 
We find that $\sim 21$\% of the cluster galaxies have clear signs of disturbances, with most of those ($\sim60$ \%)  likely being disturbed by ram pressure. The number of ram-pressure stripping candidates found (101) in A2670 is the largest to date for a single system, and while they are more common in the cluster core, they can be found even at  $> 4 \times R_{200}$, confirming cluster influence out to large radii. In support of a pre-processing scenario, most of the disturbed galaxies follow the substructures found, with the richest structures having more disturbed galaxies. 
Post-processing also seems plausible, as many galaxy-galaxy mergers are seen near the cluster core, which is not expected in relaxed clusters. In addition, there is a comparable fraction of disturbed galaxies in and outside substructures.
Overall, 
our results highlight the complex interplay of gas stripping and gravitational interactions in actively assembling clusters up to $5\times R_{200}$, motivating wide-area studies in larger cluster samples. 
\end{abstract}




\section{Introduction}
\label{sec:Introduction} 

Numerous studies have established that galaxy evolution has a strong dependence on galaxy mass as well as environment \citep[e.g.][]{Dressler1980, Goto2003, Peng2010}. The fast transition of gas-rich late-type galaxies toward passive early-types \citep[][]{Baldry2004, Baldry2006, Kormendy2012, Salim2014} implies the existence of physical mechanisms that are able to remove gas, quench the star formation and transform the morphology of galaxies. 
These can be either internal (\textit{Nature}), such as feedback from star-formation or active galactic nuclei, or external (\textit{Nurture}), through the interaction with the local or global environment. 
Several environmental mechanisms responsible for transforming galaxies have been proposed \citep[see][for a review]{BoselliGavazzi06, Cortese2021,Boselli2021a}, and they can be generally split into two broad categories:
\begin{enumerate}
    \item \textit{Gravitational} interactions between galaxies such as galaxy-galaxy \textit{mergers} \citep[e.g. ][]{Schweizer1982, MihosHernquist1994,Mihos2003}, \textit{tidal interactions} \citep[e.g. ][]{Spitzer1951,Farouki1981} and \textit{harassment} through the accumulation of several high speed encounters \citep[][]{Moore1996,moore98,Moore1999}. These kinds of interactions can distort the different components of galaxies (i.e. gas as well as stars), cause mass loss,  and eventually quench their star formation. 
    \item \textit{Hydrodynamical} interactions between the interstellar medium in galaxies and the surrounding intragroup or intracluster medium (ICM) include the shutting down of gas accretion (the supply for star formation) which results in galaxy \textit{starvation} or \textit{strangulation}  \citep[][]{Larson1980}, and \textit{ram pressure stripping} (RPS) of the interstellar medium by the ICM \citep[e.g. ][]{GunnGott1972}. 
    Because these interactions are hydrodynamical, they only affect the gas in the galaxy (not the stars). 
\end{enumerate}
All of these mechanisms are capable of removing or consuming gas, affecting star formation, and morphologically transforming galaxies. What is not yet fully understood nor quantified is the relative importance of each of these mechanisms, or the specific environments/conditions in which they are expected to be at play. 
%

Simulations and observations have shown that galaxy mergers are an effective mechanism driving the formation of  early-type galaxies \citep[e.g.][]{OkamotoNagashima2001, Diaferio2001}. However, 
galaxy mergers are relatively rare and unlikely to happen in rich clusters \citep[][]{MakinoHut1997} where velocities are too high. 
Lower-density environments with lower relative velocities between galaxies (such as groups) are more favourable places for galaxy-galaxy mergers to occur. 
In a similar way, the removal of matter from galactic halos due to tidal interactions is quite efficient 
 \citep[e.g. ][]{Richstone1976, Merritt1983, Icke1985}. It is expected that the frequency of these interactions is higher in clusters \citep[][]{BinneyTremaine1987}, 
although the high relative velocities of cluster galaxies make the encounters rather short, only allowing for longer-term harassment,  
which can also cause significant mass loss \citep{Rhee2017}.

On the other hand, hydrodynamical effects have been shown to play a crucial role in removing gas from galaxies, facilitating their later quenching. In particular, RPS can remove a significant amount of neutral gas from cluster galaxies early on during the first infall into the cluster \citep{Chung2009,Jaffe2015}, particularly for radially infalling galaxies \citep{Jaffe2018, Jaffe2018_erratum}. 
The gas stripping can sometimes lead to star-formation in the stripped gas, giving rise to the spectacular "jellyfish" galaxies observed in Optical/UV wavelengths \citep{Smith2010, Ebeling2014,Fumagalli2014,Poggianti2016,Poggianti2017a}. After the stripping phase, the galaxies are expected to become quenched \citep{Jaffe2016, Vulcani2020}.
Recent studies have shown that RPS is among the dominant mechanisms transforming galaxies in clusters, as a significant fraction  of the star-forming galaxies inside clusters have "jellyfish" morphology \citep[up to $\sim$35\%][]{Vulcani2022}. 
Milder and less frequent cases of RPS galaxies have also been found in groups and perhaps even filaments \citep{Roberts2021b, Vulcani2021, Kolku2022}, although the mechanisms at play are not always easy to separate, even with 3D data. 

Adding to the complexity of trying to separate environmental effects on galaxies, the hierarchical scenario of structure formation implies that clusters continue to grow through the accretion of smaller structures. Galaxies thus evolve within an evolving environment, transitioning from lower to higher density environments such as voids, filaments, and groups before eventually falling into clusters.  This implies that galaxies can be "\textit{pre-processed}"  (i.e. start to quench and transform) in some of these intermediate environments before they reach the cluster \citep{Fujita2004}.  Simulations indeed show that $\sim$40 per cent of cluster galaxies today 
were previously in a group \citep[][]{McGee2009,Pallero2019,Pallero2022}, highlighting the need to take pre-processing into account. 
Supporting the pre-processing scenario, observational studies have reported a large fraction of galaxies undergoing transformations (e.g. starburst) or already passive galaxies in groups in and around clusters \citep{Dressler2004, Cortese2006a,Hou2014, Boselli2014, Jaffe2016}. In addition, the fraction of star-forming galaxies in the field is significantly higher than that in cluster outskirts \citep[at $\sim 3\times R_{200}$;][]{Haines2015, Bianconi2018}, which indicates galaxies are processed in intermediate environments prior becoming cluster members \citep[see also recent work by][extending to $5\times R_{200}$]{Lopes2024}. What has not been cleared yet is what physical mechanisms are responsible for such pre-processing. 

In addition to the accretion of galaxies from filaments and groups, clusters can also grow via major mergers with other clusters. Approximately 10-20\% of low-redshift clusters show evidence of having had a major merger in the past \citep[][]{Katayama2003, Sanderson2009, Hudson2010}. There are not too many observational studies on the post-processing of galaxies in cluster mergers and the ones that have attempted it often have contradicting conclusions. Most studies find enhanced nuclear activity and star formation in cluster mergers \citep{Hwang2009, Ma2010, Stroe2014}, while some find suppressed star formation \citep{Pranger2014, Shim2011, Mansheim2017}. Moreover, observations and simulations have found evidence of increased RPS or gravitational interactions in galaxies within cluster mergers \citep{Owers2012, Rawle2014, McPartland2016, Ebeling2019, Fujita1999} and even cluster-group mergers \citep{Vijayaraghavan2013}. 
An interesting example supporting post-processing is the extremely high number of jellyfish galaxy candidates (70) in the A901/2 quadruple cluster system undergoing a merger \citep{RomanOliveira2019, Ruggiero2019}. However, a recent study attempting to quantify the fraction of jellyfish galaxies in a wide range of cluster dynamical stages found no clear evidence for an enhancement of RPS in cluster mergers \citep[][]{Lourenco2023, Kleiner2014}. Furthermore \citet{Kelkar2023} recently attempted to compare the colour and morphology of galaxies in a sample of merging vs. relaxed clusters and did not find a significant difference. They conclude that their finding is due to the longer timescale involved in the changes in global colours and morphologies of galaxies relative to the cluster mergers. They hypothesize that the lack of difference could be due to the longer timescale involved in the changing the global colours and morphologies of galaxies relative to the cluster mergers. 

In general, most observational studies of galaxy evolution in cluster regions are limited to the virial region or in some cases extend at most to 2-3$\times R_{200}$. The infall regions of clusters however are key for understanding galaxy evolution as they connect the cluster to the surrounding larger-scale structure. In fact, filaments and infalling groups are more easily detected in cluster outskirts \citep{WestBothum1990, Hou2012, Jaffe2013, Dressler2013}. Moreover, 
hydrodynamical simulations of galaxies around massive clusters have demonstrated that environmental effects such as RPS could play a role in depleting both hot and cold gas in cluster galaxies out to $\sim 5\times R_{200}$ 
\citep[][]{Bahe2013}. 

With the aim of better understanding the role of pre- and post-processing of galaxies and the relative role of the different environmental mechanisms at play, in this paper we present a detailed study of galaxies in the massive and interacting cluster Abell 2670 (A2670) and its large-scale surroundings out to $5 \times R_{200}$.
The interacting nature of the cluster offers the possibility to study post-processing while the large area studied additionally allows accounting for pre-processing. 
Our strategy is simple: we use optical imaging to identify morphological features in the cluster galaxies indicative of gravitational or hydrodynamical interactions and correlate them with their environment, which we characterize through a careful substructure analysis able to distinguish galaxy groups and other structures present in the field.

The outline of this paper is as follows: in section \ref{sec:Sec2} we summarize the main properties of A2670; in section \ref{sec:Sec3} we describe the photometric and spectroscopic optical datasets utilized to study the cluster and its galaxies; in section \ref{sec:Sec4} we identify cluster members in the photometric and spectroscopic samples; in section \ref{sec:Sec5} we perform different (2D and 3D) substructure analyses and compare the results between all the methods to robustly identify substructures in and around the cluster; in section \ref{sec:Sec6} we perform a visual inspection of the cluster galaxies and identify disturbed galaxies, split into main categories: gravitational and hydrodynamical (RPS) interactions. We then study the incidence of disturbed galaxies as a function of their local and global environment; in section \ref{sec:disc} we discuss results and compare with the literature, and finally, in section \ref{sec:Sec7}, we summarize the main results and we show conclusions. 

In this paper we adopt a $\Lambda$CDM cosmology, with a Hubble constant $H_{0} = 70$ $\text{km}$ $\text{s}^{-1}$ $\text{Mpc}^{-1}$, present matter density of $\Omega_{m} = 0.3$, and dark energy density $\Omega_{\Lambda}=0.7$.

\section{The interacting cluster Abell 2670}
\label{sec:Sec2}

\begin{figure}          
    \includegraphics[width=\columnwidth]{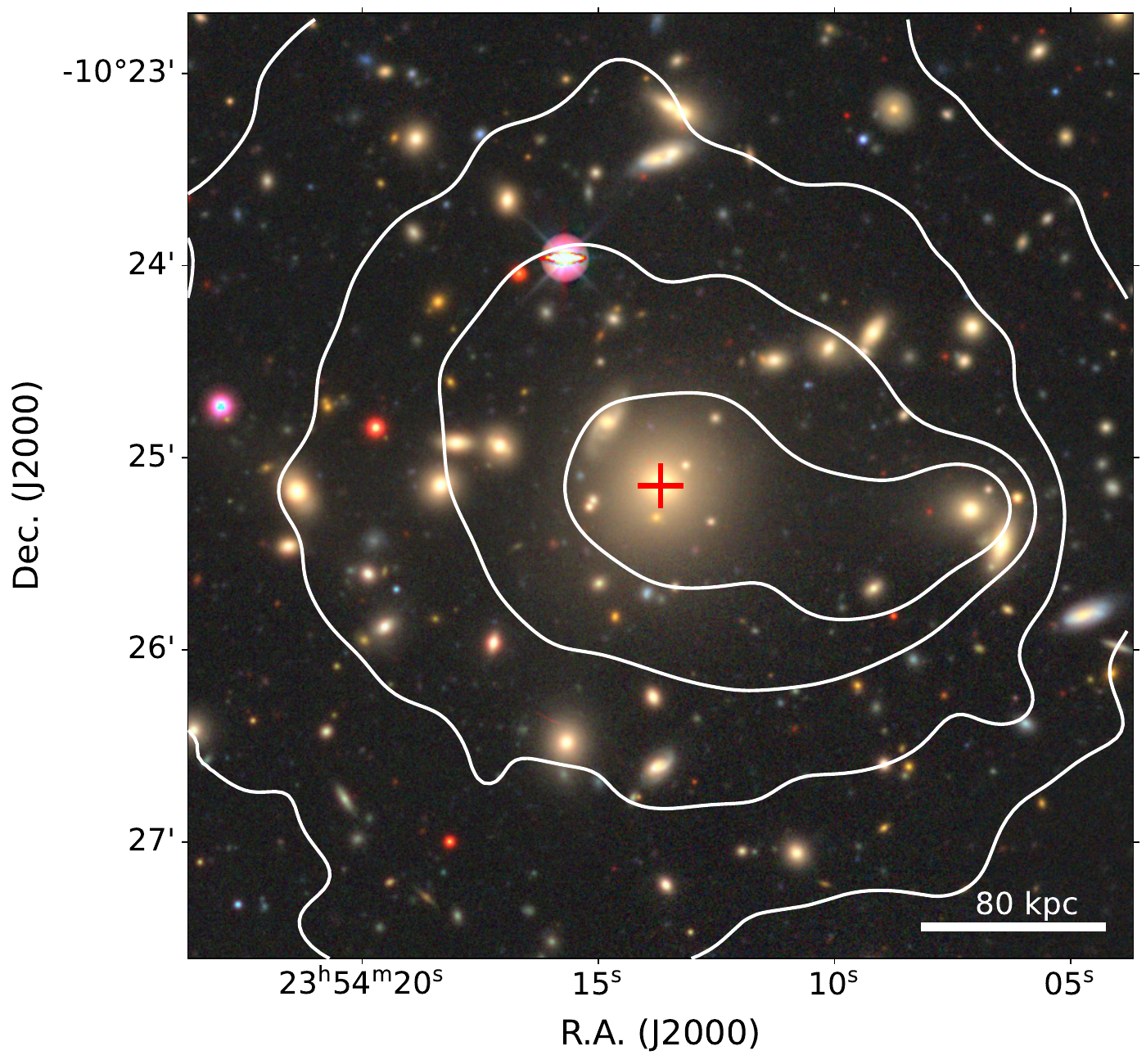}
    \caption{X-ray morphology contours from Chandra on top of colour Legacy Survey image of the central region of the cluster. The BCG is marked with a red cross. The FOV is about 5\arcmin $\times$ 5\arcmin. A scale of 80\,kpc can be visualized lower right.} 
    \label{fig:Fig1}
\end{figure}

This paper focuses on A2670, an interacting cluster at z = 0.0763, with a $\sigma_{200}$ = 919$ \pm 46$, a total mass of $M_{200}$ = $8.5 \pm 1.2~(10^{14} \mathrm{M_{\odot}})$ and a size of $R_{200}$ = $1.91 \pm 0.1$\,Mpc \citep[][]{Sifon2015a}. Previous studies analyzing the X-ray and optical emission in the central regions of the cluster show clear indications that this system is in some form of dynamical interaction along the plane of the sky \citep[][]{Hobbs1997, LopezAlfaro2022}. Although the X-ray peak is  coincident with the Brightest Cluster Galaxy (BCG)\footnote{With coordinates R.A. (J2000) = $23^{\mathrm{h}} 54^{\mathrm{m}} 13^{\mathrm{s}}.7$, Dec. (J2000) =  $ -10^{\mathrm{o}} 25' 9.2''$}, a comet-like structure in X-rays was found around one of the brightest (cD) galaxies in the central region, with a cold front at the leading edge of the structure \citep[][See also Fig. \ref{fig:Fig1}]{Fujita2006}. The mass and morphology of the X-ray gas in the comet-like structure are suggestive of the presence of a group that is being stripped as it crosses the dense ICM very near the cluster core. 

In terms of the galaxy populations in A2670, a large number of elliptical galaxies with faint features indicative of past interactions (post-mergers)  were found by \citet{Sheen2012} near the cluster centre. 
The authors concluded that these probably merged in a previous group environment prior to entering the cluster. In addition, an intriguing and unique case of ram-pressure stripping in an elliptical galaxy with blue "tadpoles" associated with it was reported by \citet{Sheen2017}. Furthermore, \citet[][]{LopezAlfaro2022} presents HI observations of this cluster out to $\sim 2\times R_{200}$ along NE-SW direction, finding three minor groups running along that direction, with strong pre-processing (HI-disturbed in this case) in the SW group.

\section{Data}
\label{sec:Sec3}

A novel and key aspect of our study is the wide-area explored around the cluster under study (A2670). To reach the far outskirts of the cluster (5$\times R_{200}$) we built a large photometric and spectroscopic catalogue in a square area of $3.6^{\circ} \times 3.6^{\circ}$ centred on the BCG. The data is 
described in the following.


\begin{figure}
\includegraphics[width=0.935\columnwidth]{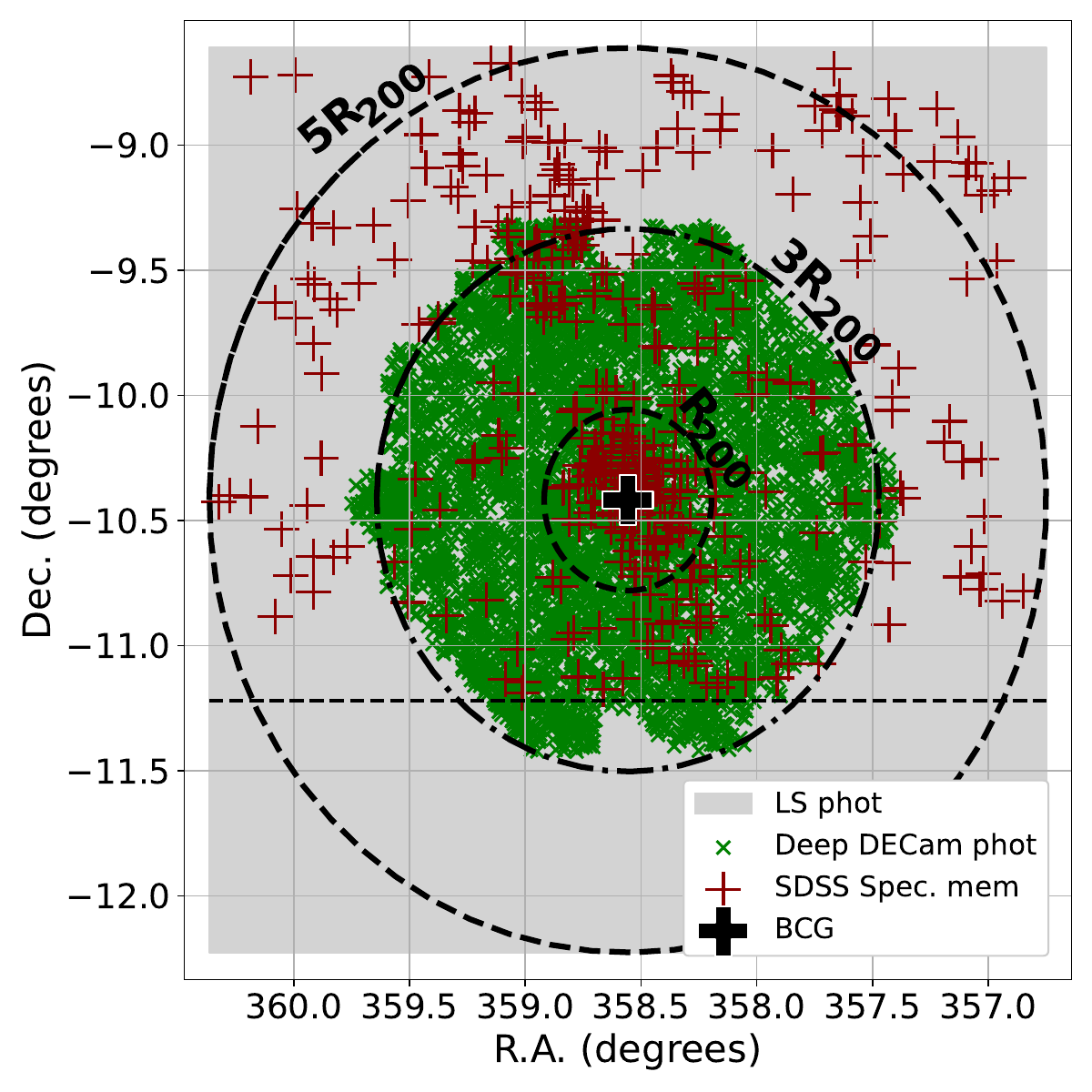}
\includegraphics[width=0.935\columnwidth]{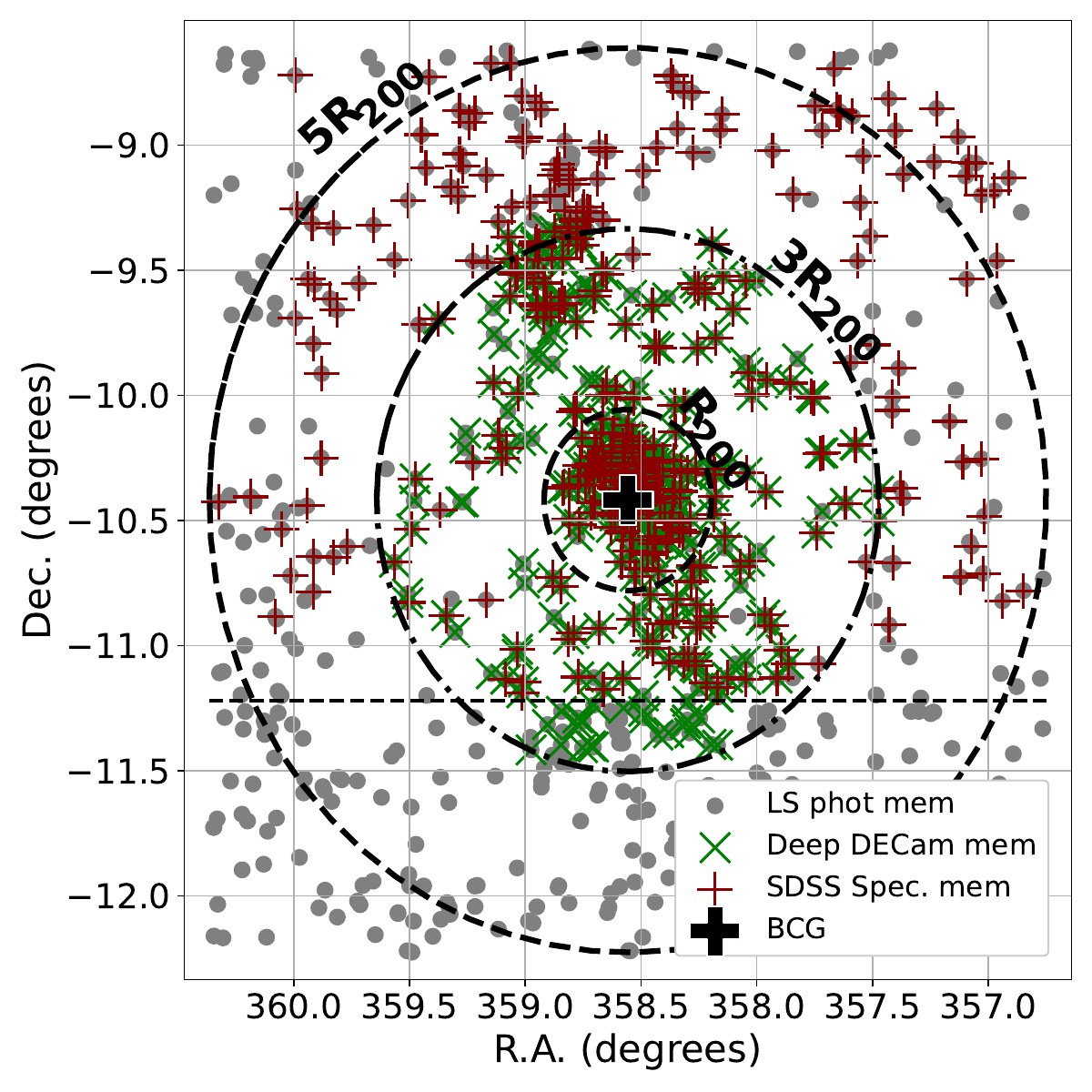}
\caption{\textbf{Top:} Distribution of the 628,902 objects from LS DR9 (smooth uniform grey pattern) and the 6,203 sources in common from our Deep DECam imaging (green) in the sky. The confirmed 405 members from available SDSS spectroscopic (red crosses) are overplotted. The dashed circles correspond to radii of 1, 3 and 5 $\times$ $R_{200}$, centred on the BCG (black cross). The black dashed horizontal line indicates the southernmost limit of SDSS. \textbf{Bottom:} The final sample of galaxies used for the substructure analysis and the visual inspection. This sample consists of galaxies within our magnitude and colour cut (see Section \ref{sec:Sec4.2}), excluding galaxies with spectroscopy that are confirmed non-members (see Section \ref{sec:Sec4.1}). This yields a total of 843 photometric cluster members (grey), a total of 364 galaxies belonging to Deep DECam imaging (green) and 376 of which are confirmed spectroscopic cluster members (red).}
    \label{fig:Fig2}
\end{figure}

\subsection{Optical Photometry}
\label{sec:Sec3.1}

\begin{figure*}          
\includegraphics[width=0.9\textwidth]{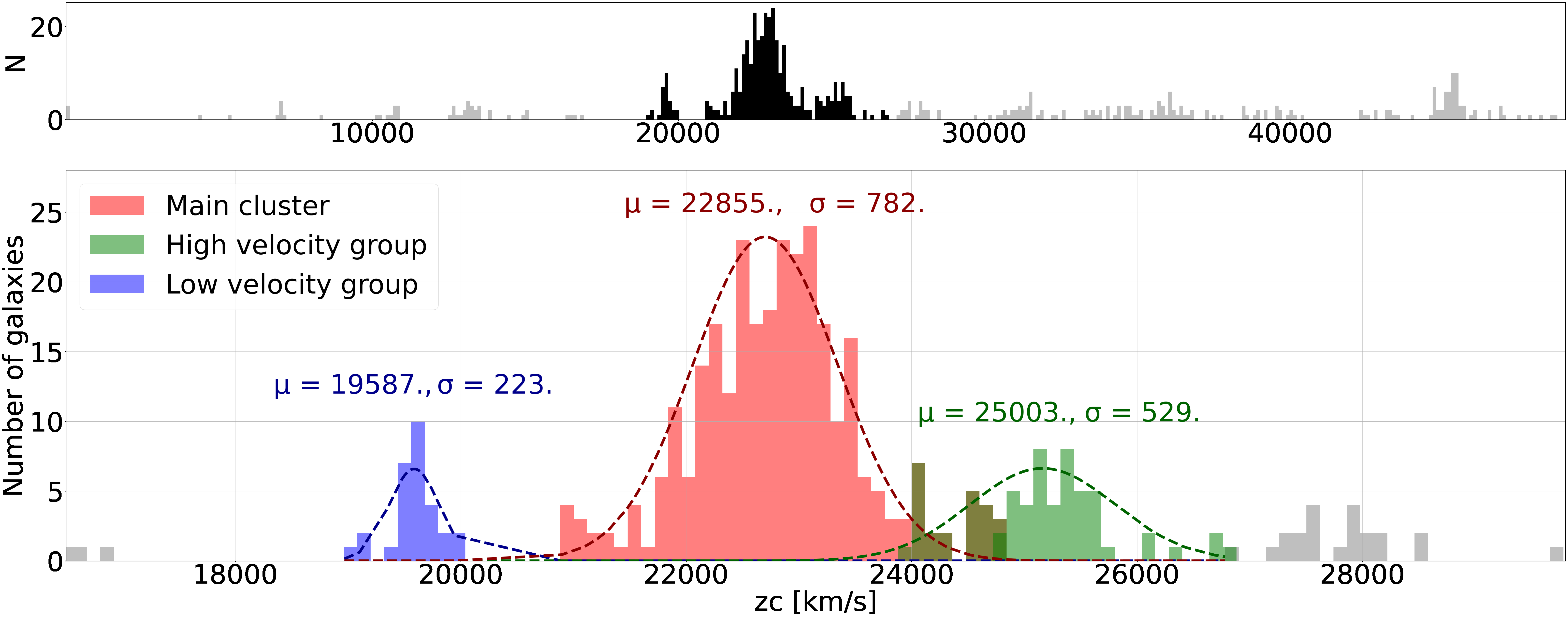}
\caption{Velocity distribution of galaxies in A2670 using SDSS spectroscopy. Three main overdensities can be easily identified. Three Gaussian were fitted and memberships were assigned for galaxies within 3-sigma from the centre of each substructure. The main cluster corresponds to the red histogram centred at $zc=22,855$ $\mathrm{km\hspace{0.1cm} s^{-1}}$ (z = 0.076), and two additional groups in the foreground and background are identified in blue and green, respectively. In this work, we consider spectroscopic cluster members all galaxies in these three structures, which translates into a redshift cut of $0.063<z<0.09$ (their distribution in the sky is shown by the red crosses in the top panel of Figure~\ref{fig:Fig2}).} 
\label{fig:Fig3}
\end{figure*}

\subsubsection{Legacy Survey}
\label{sec:Sec3.1.1}

 Our base sample consists of optical photometry from Legacy Survey \citep[LS;][]{Dey2019} public data release (DR9), obtained through the Astro Data Lab Query \footnote{\url{https://datalab.noirlab.edu/query.php}} interface.  The main tractor photometry catalogue (combined north and south region), provides both model object  type and the AB system apparent magnitudes converted from the Vega system\footnote{\url{https://wise2.ipac.caltech.edu/docs/release/allsky/expsup/sec4_4h.html}}, for 3 bands: $g$, $r$ and $z$, with their respective de-reddened magnitudes corrected by galactic extinction E(B-V). 
Duplicates as well as objects with model type = PSF (stars) were excluded. Applying these filters, the parent sample is conformed by 628,902 objects, shown in a smooth uniform grey pattern in Figure \ref{fig:Fig2} (top). 

For all the galaxies from this catalogue, we computed absolute magnitudes in r and g-band using the K-corrections of \citep[][]{Chilingarian2010, Chilingarian2012} \footnote{\url{http://kcor.sai.msu.ru/}} and a distance modulus based on the cluster redshift.

\subsubsection{Deep DECam imaging}
\label{sec:Sec3.1.2}

We also used deep optical imaging in $u$, $g$, and $r$ bands taken with the Dark Energy Camera (DECam) at the Blanco telescope in Cerro Tololo Inter-American Observatory from 19th August 2014 to 22nd August 2014 (Proposal ID: 2014B-0608, PI: Y. Jaff\'e), with 6300s, 4200s, and 9000s exposure time for the combined mosaic in $u'$, $g'$, and $r'$-band, respectively in an area of $\sim 3\,deg^{2}$, covering  $\sim 3\,R_{200}$ of the cluster. The details of the data reduction and processing can be found in Kim et al. (submitted). This data is about a magnitude deeper than LS in $g'$, and $r'$. 

After source extraction in the deep DECam images, we found 6,203 galaxies in common between the LS catalogue and the deep DECam catalogue. These are plotted as green crosses in Figure~\ref{fig:Fig2} (top). 

\subsection{Optical Spectroscopy}
\label{sec:Sec3.2}

Public spectroscopic data from Sloan Digital Sky Survey (SDSS) DR9 \citep[][]{SDSS_DR9} were also collected  for galaxies in the large area explored in this work, but these are only available for a declination $\gtrsim-11.\!\!^{\circ}2$ (dashed horizontal line in Figure~\ref{fig:Fig2}). When matching the SDSS spectroscopic data with our parent sample, we found 1685 objects with redshift, down to the SDSS spectroscopic magnitude limit of $m_{r} \lesssim 17.5$, which roughly corresponds to $M_{r} \lesssim -20$.  
The spectroscopic redshift was used to define cluster membership as discussed in the following section. 

\section{Cluster members}
\label{sec:Sec4}

\subsection{Spectroscopic cluster members}\label{sec:Sec4.1}

We use the available SDSS spectra for galaxies located in the square area defined in Section~\ref{sec:Sec3} (Fig.~\ref{fig:Fig2}) in order to define cluster membership and to identify substructures. Figure~\ref{fig:Fig3} (top) shows the velocity distribution ($zc$) for all the galaxies with spectroscopy in our sample. Three overdensities are observed in the blackened region. If we zoom in taking a velocity range covering this region (Fig \ref{fig:Fig3}, bottom), we can find a clear peak for the main cluster (red histogram). We can also identify two subgroups at low ($zc=19,587$ $\mathrm{km\hspace{0.1cm} s^{-1}}$, $\sigma = 223$ $\mathrm{km\hspace{0.1cm} s^{-1}}$) and high ($zc=25,003$ $\mathrm{km\hspace{0.1cm} s^{-1}}$, $\sigma = 529$ $\mathrm{km\hspace{0.1cm} s^{-1}}$) redshift, likely two accreting foreground/background substructures (blue and green histograms in Fig \ref{fig:Fig3}, respectively).

The departure from a single Gaussian distribution is already a sign that the cluster is not relaxed. To define cluster membership we first selected a sample of galaxies in three components/peaks visible in the redshift distribution, and then we fit three individual Gaussians, one to each of the components. We then considered galaxies within $\pm 3 \sigma$ of the mean velocity of each Gaussian to be part of each structure.
The resulting mean velocities ($\mu$) and dispersion $\sigma$ for each structure are marked in Figure~\ref{fig:Fig3}. Galaxies with velocities in the range $23,983.39 < zc < 24,806.67$ $\mathrm{km\hspace{0.1cm} s^{-1}}$ (dark overlapping histogram in the figure) could belong to either/both the high-velocity group or the main cluster.

Given our scientific goal and the complex nature of this cluster, we define a generous final cluster membership that includes the main cluster as well as its foreground/background groups, covering a velocity range starting from the lower limit for the low-velocity group up to the higher limit for the high-velocity group ($18,916.6 - 27,036.66$ $\mathrm{km\hspace{0.1cm} s^{-1}}$). Using this definition, we obtain 405 spectroscopic cluster members and 1280 non-members from the 1685 galaxies with spectra in our sample. Spectroscopic cluster members galaxies are plotted in Figure~\ref{fig:Fig2} (top) as red plus markers (spectroscopic non-members are not plotted). We computed, for reference, the velocity dispersion of the core cluster members (red histogram) within $R_{200}$, obtaining $\sigma_{200}$ = $874$ $\mathrm{km\hspace{0.1cm} s^{-1}}$.

\subsection{Photometric cluster members}
\label{sec:Sec4.2}

\begin{figure}          
    \includegraphics[width=0.96\columnwidth]{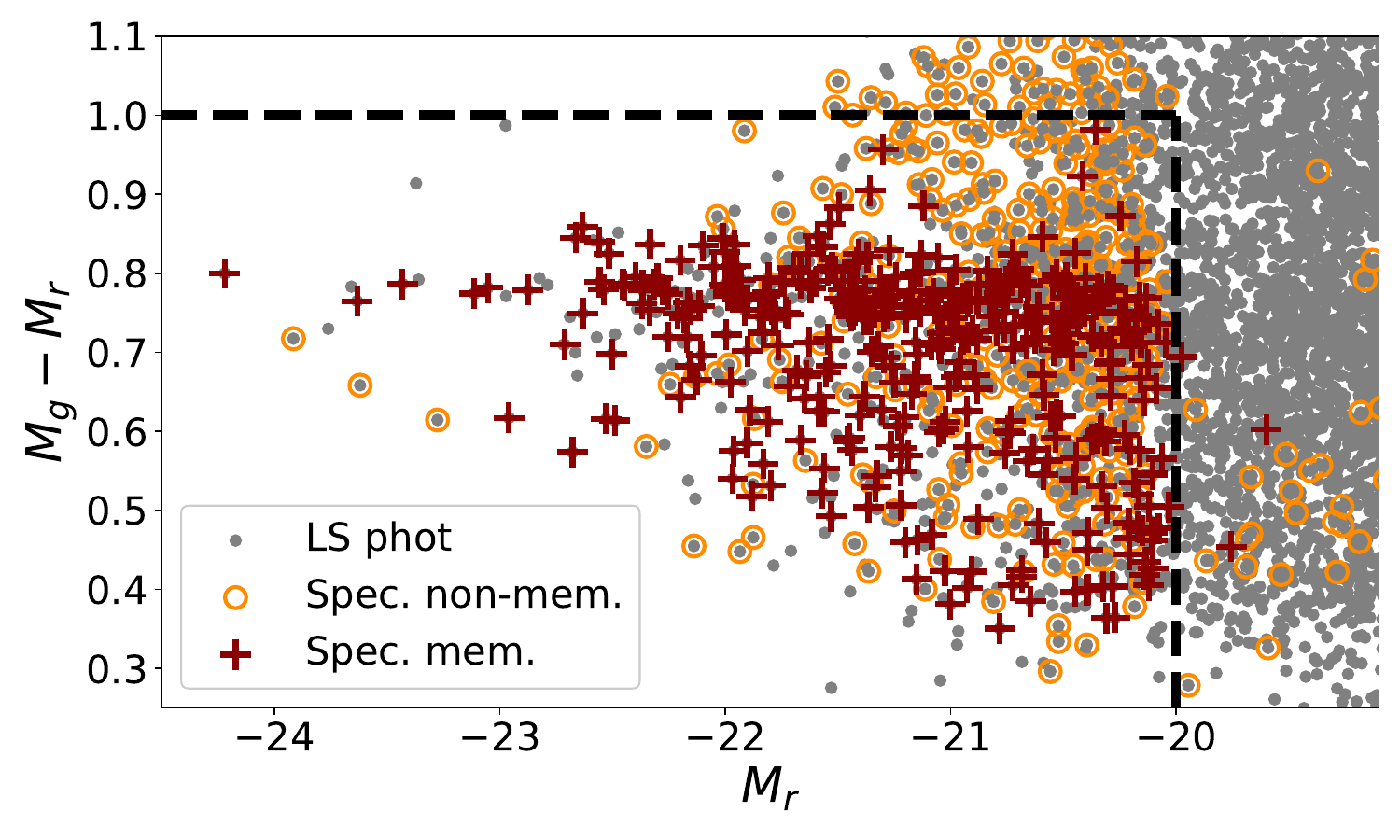}
    \includegraphics[width=0.98\columnwidth]{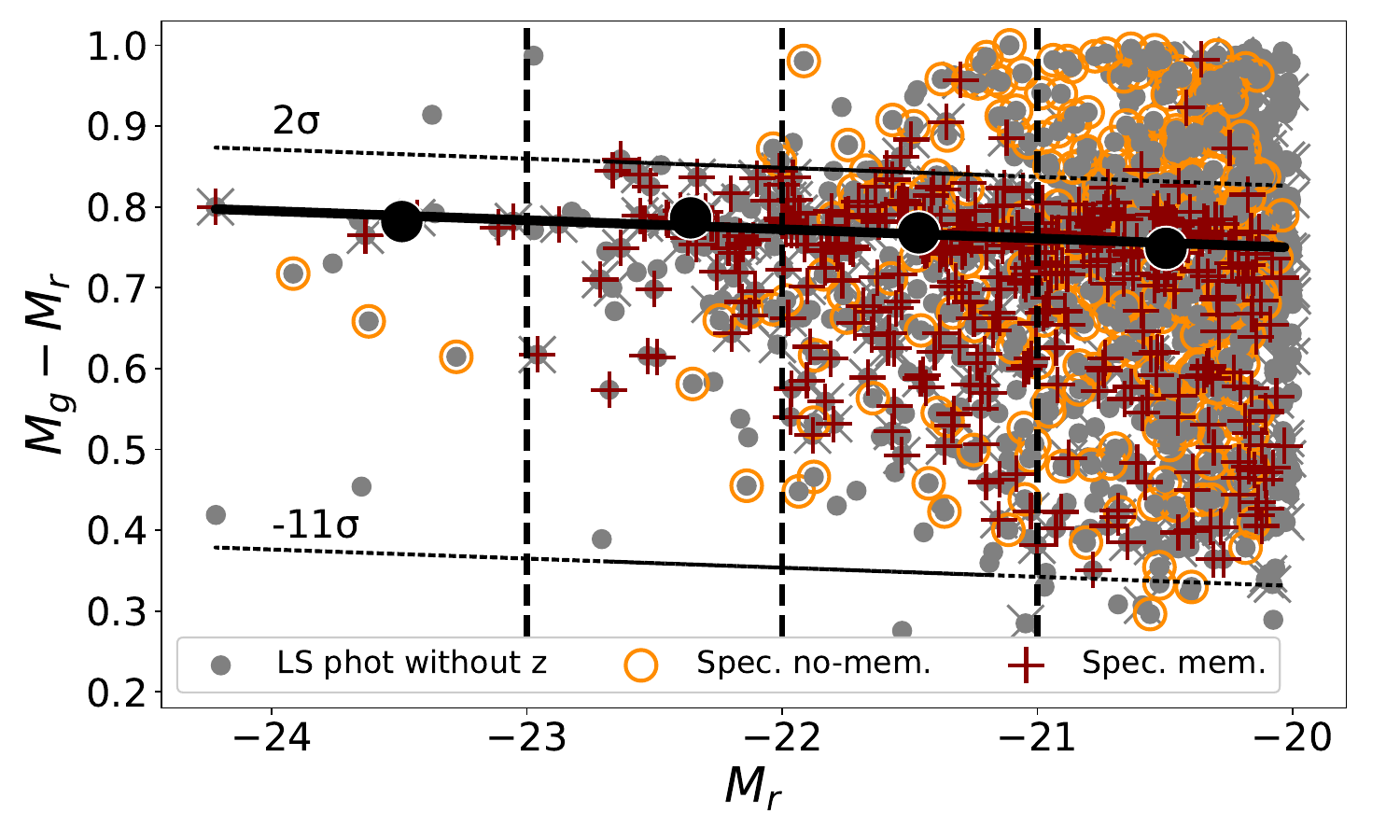}
    \includegraphics[width=1.05\columnwidth]{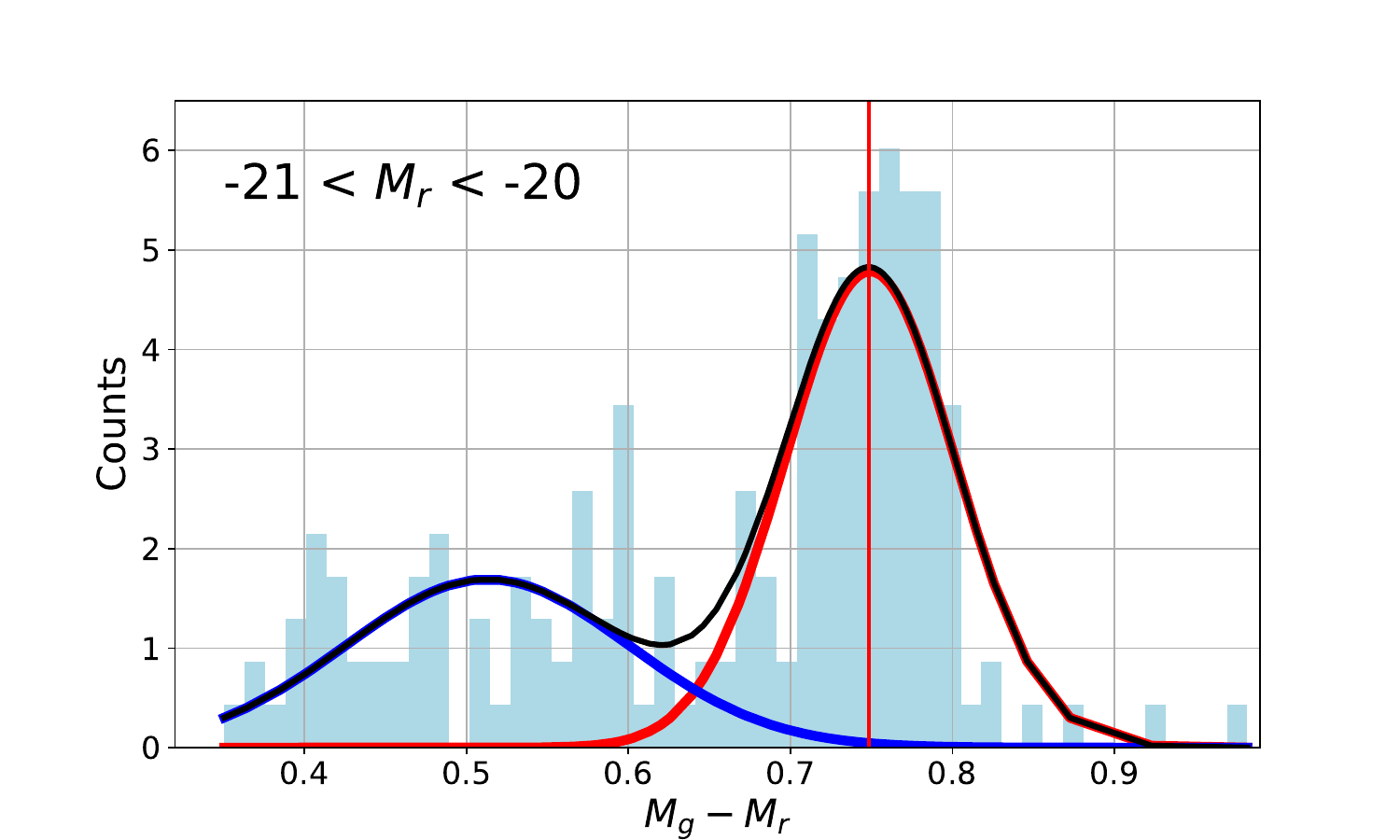}
    \caption{\textbf{Top:} Color Magnitude Diagram for the complete LS photometric sample (grey) where we highlight the spectroscopic sample both no-member (orange) and members (red) (Section \ref{sec:Sec4.1}). To maximize the probability of a given galaxy being a cluster member, we started with a reasonable cut in $M_{r} < -20$ (horizontal black dash line) following \citet[][]{Sheen2012} and also because it coincides with the spectroscopic limit from SDSS ($m_{r} \lesssim $17.5) and we considered the colour range of the confirmed spectroscopic members, which yielded $M_{g} - M_{r} < 1$ (vertical black dash line). \textbf{Middle:} The four bins across the r band magnitude. For each bin and using only the confirmed spectroscopic members, a double Gaussian to the colour distribution was fitted. The peak value of the redder gaussian of each bin and the measured absolute magnitude corresponding to this peak are plotted as large filled circles and then the line was fitted which corresponds to the red sequence (dark black line). \textbf{Bottom:} An example of the double gaussian fit performed on the fainter magnitude bin ($M_{r}$ from -21 to -20 mag). The vertical red line identifies the peak colour of the red galaxy population in that magnitude bin.} 
    \label{fig:Fig4}
\end{figure}

To increase the number of galaxies studied and to reach 5$\times R_{200}$ homogeneously, 
we assemble magnitude-limited sample of galaxies that are likely part of the cluster using photometric and spectroscopic information. To define the photometric memberships, we use the LS photometry which is homogeneously available throughout the cluster, and SDSS spectroscopy to reject known non-members where available. 

We adopt a cut in $M_{r} < -20$ following \citet[][]{Sheen2012}. This cut allows for reasonable visual classification of the galaxies and at the same time, it coincides with the magnitude limit reached by SDSS spectroscopy ($m_{r} \lesssim $ 17.5). We further consider colour cuts to maximize the probability of a given galaxy being a cluster member. We first identified the colour range of the confirmed spectroscopic members, which yielded $0.3 < M_{g} - M_{r} < 1$ (the dashed box in Fig. \ref{fig:Fig4} upper). Then, using the confirmed spectroscopic members from the SDSS catalogue (subsection \ref{sec:Sec4.1}), we defined the \textit{red sequence} and \textit{blue cloud} galaxies within this box by fitting a double Gaussian to the colour distributions in 4 bins of $M_{r}$ (delimited by the vertical lines in the middle panel of Figure~\ref{fig:Fig4}). 
For the brightest bin, given the lack of blue galaxies, we only fitted a single Gaussian. In the bottom panel of Figure~\ref{fig:Fig4}, we show as an example the double Gaussian fit performed to the fainter magnitude bin ($M_{r}$  from -21 to -20 mag). The vertical red line identifies the peak colour of the red galaxy population in that magnitude bin. Once this was done for all magnitude bins, we took the peak value of the redder Gaussian to fit a red sequence, using the mean of $M_{r}$ of all spectroscopic member galaxies in each bin (see large filled black circles in the middle panel of the figure). The resulting fit is shown as a dark black line in the middle panel of the figure. 


Finally, we considered photometric cluster members those galaxies above our magnitude limit. In order to avoid significant contamination from outliers in our selection (i.e. spectroscopic non-members, plotted as orange circles in the middle panel Fig.~\ref{fig:Fig4}) without sacrificing possible cluster members, we further make a colour cut at $-11\sigma$ and at $+2\sigma$ from the red sequence fit (dotted lines in the middle panel of Figure~\ref{fig:Fig4}). 

We can assess the purity of the photometric membership selection in the region where we have SDSS spectroscopy. From the colour-magnitude selection we have 1053 galaxies with photometry, for which we have spectra for 586 (56\%) of them. Of those, a total of 376 are confirmed spectroscopic members. 
If we only consider the spectroscopic sample, the expected contamination of interlopers from a purely colour-magnitude selection is 36\%. But in our selection we do not just consider colour but we also remove spectroscopically confirmed non-members (orange circles in Fig.~\ref{fig:Fig4}) when spectra is available, i.e. for  56\% of the photometric sample, so our final contamination of interlopers will reduce to $\sim$ 16\% in the region covered by SDSS and its inevitably higher in the southernmost region. 
Relative to the spectroscopic sample, the photometric sample  increases the number of galaxies studied by 44\%  and covers 5$\times R_{200}$ homogeneously. 
%


In summary, there are 843 photometric cluster members the extended region around A2670 reaching $5 \times R_{200}$. From that parent sample 376 galaxies are spectroscopically confirmed  cluster members (Fig.~\ref{fig:Fig2}, bottom panel), and 364 have deep DECam imaging (green crosses; see Subsec. \ref{sec:Sec3.1.2}).


\section{Cluster Substructure analysis}
\label{sec:Sec5} 


In Section~\ref{sec:Sec4.1}, we found preliminary evidence through the galaxy velocity distribution that supports the merger scenario presented in the literature (see Section~\ref{sec:Sec2}). In this section, we characterize the environment in and around A2670 out to its outskirts through the identification of substructures using two different methods.

\subsection{Dressler-Shectman's test}
\label{sec:Sec5.1}

The Dressler-Schetman \citep[DS,][]{Dressler1988} test is a statistical test widely used in astronomy to establish whether a cluster is disturbed and to highlight galaxy substructures. In the case of a galaxy cluster, the DS analysis will highlight smaller galaxy groups clustered in the sky by measuring their deviations in velocity from the velocity distribution of the galaxies in the main cluster.

In short, the test consists of the following steps: 
We identify the $N_{nn}$ nearest neighbours of each galaxy. Then, for each i-group of $N_{nn} + 1$ galaxies, we compare the \textit{local} mean velocity $\overline{v}^{i}_{local}$ and dispersion $\sigma^{i}_{local}$ with the \textit{global} mean velocity $\overline{v}_{cl}$ and dispersion $\sigma_{cl}$ of the cluster. This is done by computing the deviation $\delta_{i}$ parameter, defined as: 

\begin{equation}
    \delta_{i}^{2} = \left( \dfrac{N_{nn} + 1}{\sigma^{2}_{cl}} \right)
    [(\overline{v}^{i}_{local} - \overline{v}_{cl})^{2} 
    + (\sigma^{i}_{local} - \sigma_{cl})^{2}]
    \label{eq:delta_i}
\end{equation}

The larger the value of $\delta_{i}$, the larger the deviation of the local parameters with respect to the global ones.

We applied the DS test to all confirmed spectroscopic members from the photometric cluster members sample. 
In order to highlight the infalling structures, the cluster parameters $\overline{v}_{cl}$ and $\sigma_{cl}$ were determined considering only members from the main cluster, that are within $R_{200}$ (see Section~\ref{sec:Sec4.1}), using
$N_{nn} = 10$ following similar studies \citep[e.g.][]{Jaffe2013}.

\begin{figure*}          
\begin{center}
\includegraphics[width=0.865\textwidth]{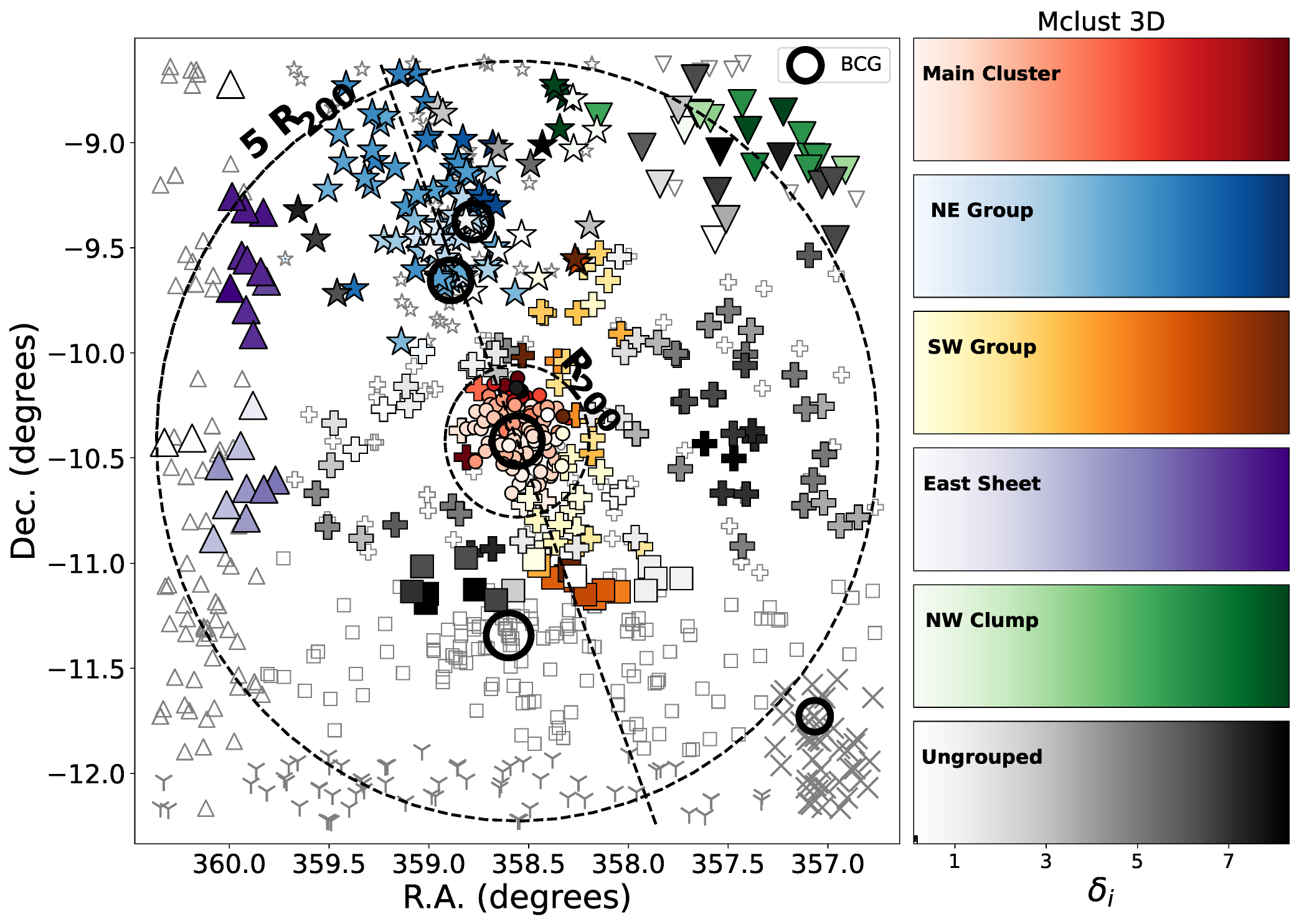}
\includegraphics[width=0.865\textwidth]{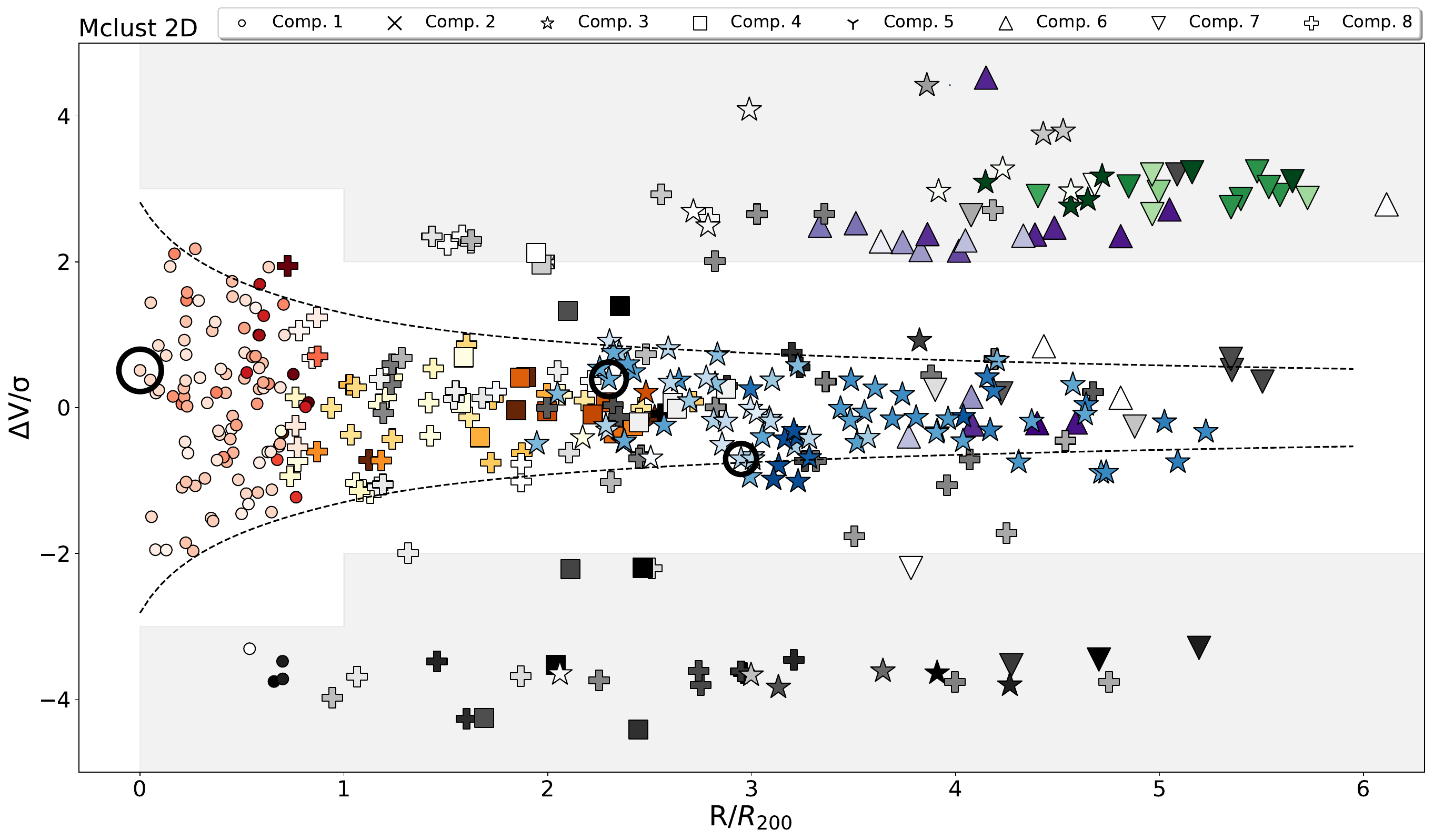}
\end{center}
\caption{Distribution of the photometric cluster members in the sky (top) and the spectroscopic cluster members in position vs. velocity phase-space (bottom). The substructures identified by \textit{mclust} 2D and 3D are highlighted with different symbols and colours respectively, while the darkness of the colour increases with increasing $\delta_i$ (from the DS analysis), as labelled.  Note that because not all photometric members have a spectroscopic redshift, some galaxies that belong to structures found by \textit{mclust} 2D are not in the 3D sample (i.e. the smaller grey open symbols, which are absent in the bottom plot). The inner and outer circles in the top panel correspond to 1 and $5 R_{200}$, respectively. The 5 brightest galaxies are also plotted as solid black open circles, with the sizes reflecting their relative magnitudes (in the phase space only those with redshifts are plotted). The dashed line in the top panel shows the merger axis of the cluster, which connects the Main cluster (or Comp. 1) with the NE Group (Comp. 3). The dashed curves in the bottom panel correspond to the escape velocity in a NFW halo \citep[][]{NFW96} with the concentration parameter equal six, similar to \citet[][]{Jaffe2015}. In the same panel we can see a shaded region corresponding to the unbound galaxies that are excluded.}
\label{fig:Fig5}
\end{figure*}

In Figure~\ref{fig:Fig5}, we show the distribution of galaxies in the plane of the sky and in projected position vs. velocity (phase-space), and highlight galaxies with large $\delta_{i}$ using darker colours (the different symbols and colours are explained in Sec.~\ref{sec:Sec5.2.1} and \ref{sec:Sec5.2.2}, respectively), where the probability that a galaxy belongs to a substructure increases with $\delta_{i}$. The fact that galaxies with large $\delta_{i}$ tend to cluster together in the sky is  a strong indication of the presence of well-defined substructures.

To confirm and quantify the disturbed nature of the cluster, we performed the two standard statistical DS tests:

\begin{enumerate}

\item \textbf{Critical value:} A cumulative deviation $\Delta$ is computed as the sum of $\delta_{i}$: 

\begin{equation}
    \Delta = \sum_{i} \delta_{i}
    \label{eq:Delta}
\end{equation}

Values of $\Delta/N_{mem} > 1$ indicate strong evidence of substructures, where $N_{mem}$ is the number of cluster members.

\item \textbf{P-value}: 
To further test the statistical robustness of the $\Delta$-value test, it is possible to randomly shuffle  the observed radial velocities of each  cluster member and reassign them to a new galaxy of the sample (i.e. Monte Carlo shuffling), in order to compute 'shuffled' $\Delta$-values. Comparing these $\Delta_{shuffle}$ values with our $\Delta_{obs}$ computed from Eq. \eqref{eq:Delta}, we can obtain a P-value defined as 

\begin{equation}
    P = \sum ( \Delta_{shuffle} > \Delta_{obs}/N_{shuffle})
    \label{eq:Pvalue}
\end{equation}

Where $N_{shuffle}$ is the number of Monte Carlo shuffles performed, typically around 5000. Values of P $<$ 0.01 provide a robust constraint on the presence of substructures in the clusters. 

\end{enumerate}

We obtain $\Delta/N_{mem} = 2.83$ and P $<$ 0.001 after applying the DS test in A2670. We confidently conclude that the cluster hosts significant substructures. One of the disadvantages of the DS test is not being able to identify which galaxies are members of the substructures. To this aim, we applied a second test described in the following. 

\subsection{Mclust}
\label{sec:Sec5.2}

\textit{Mclust} \citep[][]{Scrucca2016} is an R package for model-based clustering and density estimation using finite Gaussian Mixture Models (GMM). The code applies the Expectation-Maximization (EM) algorithm to estimate the parameters of GMM for clustering. It provides a wide range of covariance structures with varying shapes, volumes, and distributions of the covariance ellipses. \textit{Mclust}, by default, identifies the model that best fits the data using the Bayesian Information Criterion \citep[BIC,][]{KassRaftery1995}. The code also allows the user to visualise the clustering results and assess the model fit. \textit{Mclust} has been widely used in astronomy for substructure analyses \citep[e.g.][]{Einasto2012a,Einasto2012b,MonteiroOliveira2020,MonteiroOliveira2022,Lourenco2020}

\subsubsection{Mclust 2D}
\label{sec:Sec5.2.1}


We first apply \textit{mclust} in 2D to the spatial distribution of all photometric cluster members (defined in Sec~\ref{sec:Sec4.2}) using their coordinates as inputs (R.A. and Dec. in degrees). When running \textit{mclust} in 2D, different two dimensional gaussian mixture models are fitted across the distribution of galaxies on the plane of the sky. The model that converged to the highest BIC found a total of 8 components, that vary in volume and shape between each other \citep[VVI model, see table 3 in][]{Scrucca2016}.

In Figure~\ref{fig:Fig5}, the eight components are highlighted with eight different symbols and labelled with numbers increasing inversely with the galaxy density: Component 1 (Comp.~1) corresponds to the densest substructure and Component 8 (Comp.~8) the lower density one. For reference, the five brightest galaxies are also plotted as larger filled circles, with the sizes reflecting their relative magnitudes. These galaxies coincide with some of the substructures found and have similar magnitudes (difference of 0.56-0.86 mags) which according to \citep{Raouf2019} is indicative of cluster interaction. 

The densest structure (Comp.~1, filled circles in Fig. \ref{fig:Fig5}) is a group of 146 galaxies ($>17\%$ of all the photometric members) located at the centre of A2670. This structure coincides spatially with the X-ray emission (see Figure~\ref{fig:Fig1}) and the BCG (solid open circle at the centre of Fig~\ref{fig:Fig5}), clearly indicating this component corresponds to the main cluster (A2670). The second densest component is located southwestward of the main cluster at $> 5 R_{200}$ (Comp.~2, x symbols in Fig. \ref{fig:Fig5}), where we see a compact clump with 46 galaxies distributed around the 5th brightest galaxy in the field. It is unclear if this group extends further as it is at the edge of the inspected field. A third high-density component with 144 galaxies was detected northeastward of the main cluster (Comp.~3, stars in Fig. \ref{fig:Fig5}). This group has the 3rd and 4th brightest galaxies in the field and is comparable in number of galaxies to the main cluster, but it is less dense, and extends from $\sim$2 to $\sim5 R_{200}$. Based on their proximity in position and velocity (see both panels in Figure~\ref{fig:Fig5}) and the fact their BCGs have a similar magnitude (with a magnitude gap of $\sim 0.6$ mags),  it is likely that Comp.~3 and Comp.~1 are in the process of merging with each other. The fourth densest structure (Comp~4, squares in Figure~\ref{fig:Fig5}) is the most  numerous one, containing over 19\% of the galaxies. This loose structure is located southward of the main cluster, has the second brightest galaxy in the field, and extends from $\sim$ 1.5 to $\sim5 R_{200}$. 
Further south another very loose and extended (along R.A.) component is detected at the edge of the field (Comp.~5, propellers in Fig. \ref{fig:Fig5}) with 61 galaxies. 
Towards the east, there is also an extended (along DEC.) structure of 96 galaxies (Comp.~6, triangles in Fig. \ref{fig:Fig5}) and at the north-west corner a low-density but well-separated group of 36 galaxies (Comp.~7, inverted triangles in Fig. \ref{fig:Fig5}). Both of these structures have high relative velocities ($\sim 3\sigma$) with respect to the main cluster, as shown in the bottom panel of Figure~\ref{fig:Fig5}. 
Finally, the analysis found 152 galaxies that do not seem to be grouped (Comp.~8, crosses in Fig. \ref{fig:Fig5}). This component is indeed the least reliable "group" because the second-best model of our analysis (i.e. that with the second-highest BIC, which was only marginally smaller) excluded this structure.

\subsubsection{Mclust 3D}
\label{sec:Sec5.2.2}

We add spectroscopic information on the line of sight velocity to perform a three-dimensional (3D) analysis for the confirmed spectroscopic members, which is more robust but limits the sample in numbers and also in the area covered (See Figure~\ref{fig:Fig2} and section \ref{sec:Sec4.1}). 
In this case \textit{mclust} uses R.A., Dec., and redshift information of the galaxies across the cluster to fit different 3D gaussian mixture models. 
The model with the highest BIC found by \textit{mclust} 3D  is the same as for \textit{mclust} 2D, but this time, finding only 6 different components. 

In Figure \ref{fig:Fig5}, the six components are highlighted with six different colours and labelled according to their position on the plane of the sky: The Main Cluster (red in Fig. \ref{fig:Fig5}) is the densest and dominant structure at the centre, with a total of 102 galaxies ($> 25\%$ of the spectroscopic cluster members) clearly corresponds to Comp.~1 found by \textit{mclust} 2D. A second less populated group of 76 galaxies at the same mean velocity of the cluster and located $>2 R_{200}$ northeastward was labelled NE Group (blue in Fig. \ref{fig:Fig5}), and largely corresponds to Comp.~3. 
A third group with 48 galaxies at the same velocity of the main cluster, and located mostly towards southwest, but also partially extending to the northwest was labelled as SW Group (orange in Fig. \ref{fig:Fig5}).  
This group includes galaxies from Comp.~4 and Comp.~8 from the 2D analysis. 
A pair of poorly populated but clearly separated substructures at high relative velocities were found towards the east and towards the northwest, and were called East Sheet (purple in Fig. \ref{fig:Fig5}) and NW Clump (green in Fig. \ref{fig:Fig5}) respectively. These largely correspond to Comp.~6 and~7. 
Finally, we have a sixth "Ungrouped" component (grey in Fig. \ref{fig:Fig5}), that spans a broad area and velocity range and hence it's not well defined. More than 25\% of the spectroscopic cluster members belong to this component, with a total of 108 galaxies. 

For this analysis, only the SW group seems to be the least reliable "group", since its composition shows a strong uncertainty in comparison with the other components. In fact, the second-best model detects a seventh component composed of some galaxies that belong to SW Group, mainly those that are extended northward. In addition, only some of the southern galaxies in this group seem to be part of a similar group found by \textit{mclust} 2D in the same direction (Comp.~4).

\subsection{Comparison of the different substructure analysis}
\label{sec:Sec5.3}

In general, there is a good agreement between the structures found by the 3D and 2D analysis as they both robustly find the main cluster (Comp.~1), a large group towards the NE (Comp.~3) which are likely merging with each other, and thus define the merger axis of the system (dotted line in the top panel of Figure~\ref{fig:Fig5}). There are also two clear lower-density high-velocity substructures which are found by both analysis: The East Sheet (Comp.~6) and the NW Clump (Comp.~7). However, the southern part is not subject to comparison between the 2D and 3D analysis, since this region is not covered by the spectra and it is difficult to predict if the SW Group could extend beyond the limited declination (see Section \ref{sec:Sec3.2}) forming a similar southward component like at \textit{mclust} 2D (Comp.~4).

One caveat of \textit{mclust} is that a mixture of Gaussians does not necessarily describe the data distribution. Not all the galaxies in the sample belong to the group that the GMM assigned them to. Using the \textit{mclust} with the DS test allows us to validate the substructures of the different methods. 
Indeed there are higher values of $\delta_i$  in the substructures found by \textit{mclust}, as seen by the darker colours in Figure~\ref{fig:Fig5} in all of these structures except for the main cluster which was taken as reference in the DS analysis.

Finally, the bottom panel of Fig.~\ref{fig:Fig5} shows there are galaxies that are not bound to the cluster. 
Among these are galaxies from the East Sheet (Comp.~6) and the NW Clump (Comp.~7), which  match the high velocity group seen in Fig. \ref{fig:Fig3} (green histogram). 
The low-velocity group in Fig. \ref{fig:Fig3} (blue histogram) can also be seen at the bottom of the plot, as a clear sheet of galaxies located in $\Delta V/ \sigma < -3$, however, it does not have a clear 2D or 3D substructure associated with it. This sheet seems to be composed mainly by ungrouped galaxies or members of different 2D components, with high $\delta_{i}$ values from DS-test.

In the rest of the analysis, we name all galaxies with velocities close to the caustic line in the bottom panel of Fig.~\ref{fig:Fig5} as \textit{bound cluster galaxies}. Specifically, we do not consider galaxies with $|\Delta v| / \sigma > 3$ at $R/R_{200}<1$ and $|\Delta v| / \sigma > 2$ at $R/R_{200}>1$ (shaded region in phase-space diagram).

\section{Visual inspection of the cluster galaxies}
\label{sec:Sec6} 

The large amount of substructures found in Section~\ref{sec:Sec5} 
confirms that A2670 is indeed a complex merging/interacting system. In order to understand how the galaxies in this system are affected by the local and global environment, eight classifiers 
visually inspected all the photometric cluster members (see Section \ref{sec:Sec4.2}) 
to search for disturbed galaxies, including those with signs of ongoing or past interactions (e.g. mergers, post-mergers), and those with hydrodynamical effects (e.g. ram-pressure stripping)
, independently of their morphology. 
To perform the visual inspection, we made $60\arcsec \times 60\arcsec$ RGB image cutouts from LS sky browser\footnote{\url{https://www.legacysurvey.org/viewer}} for the photometric and spectroscopic cluster members. We also created cutouts fits in u-, g-, and r-band with the same size for the galaxies with deep DECam data available. In addition, we used r-band and composite u-, g-, r-color images available as well for galaxies from Deep DECam imaging catalogue (see Kim et al. submitted for more details). 

In the following, we explain the tools and methods used to classify the galaxies (Sec.~\ref{sec:Sec6.1}) and present the population of disturbed galaxies found in the cluster and substructures (Sec~\ref{sec:Sec6.2}). 

\subsection{Identification of \textbf{disturbed} galaxies}
\label{sec:Sec6.1}

We developed a custom-made tool in python to inspect and classify each galaxy \footnote{\url{https://github.com/FrancoPiraino/Interacting_Galaxy_Classification.git}}. 
The code displays several images of a given galaxy: a colour RGB composite image from LS, alongside the deep DECam g and r-band images (when available) and $ugr$-color composite stamps (Kim et al. submitted). Each classifier visually inspects the stamps to assign a classification. 
Eight authors (FP,YJ,VS,JC,AL,KK,DP,DK) independently 
classified all or a subset of the total galaxy sample, assigning for each of them, one of the following categories: 

\begin{itemize}
    \item \textbf{Merger (M):} Ongoing merger between two or more galaxies that resulting in a very disturbed system, with multiple cores, or features such as tidal tails or stellar bridges between galaxies.
    \item \textbf{Post-merger (PM)}: Typically bright elliptical galaxies with faint features around the galaxy such as shells that indicate a past gravitational interaction/merger \citep[see][]{Sheen2012, Sheen2017}. 
    \item \textbf{Ram-pressure stripping or "jellyfish" candidate (JF):} Galaxies with features indicative of gas stripping such as one-sided asymmetries, tails, star formation in one side and/or unwinding spiral arms   \citep[see][]{Poggianti2016, Bellhouse2021, Vulcani2022}. These features had to have blue colours in the RGB composite images from LS. 
    \item \textbf{Gravitational or hydrodynamical (JM):} Cases of disturbed galaxies where it is not clear whether the disturbance is gravitational (i.e. M or PM), hydrodynamical (JF), or both at the same time.
    \item  \textbf{Star (S):} Very bright point-like sources that could be stars.
    \item \textbf{Broken (B):} The classification was not possible due to artefacts such as saturation from nearby stars. 
    \item \textbf{Nothing (N):} Normal galaxy of any type (spiral, lenticular, elliptical or irregular) with no clear sign of gravitational or hydrodynamical disturbances.
\end{itemize}

\begin{figure*}         
    \begin{center}
    \includegraphics[width=0.9\textwidth]{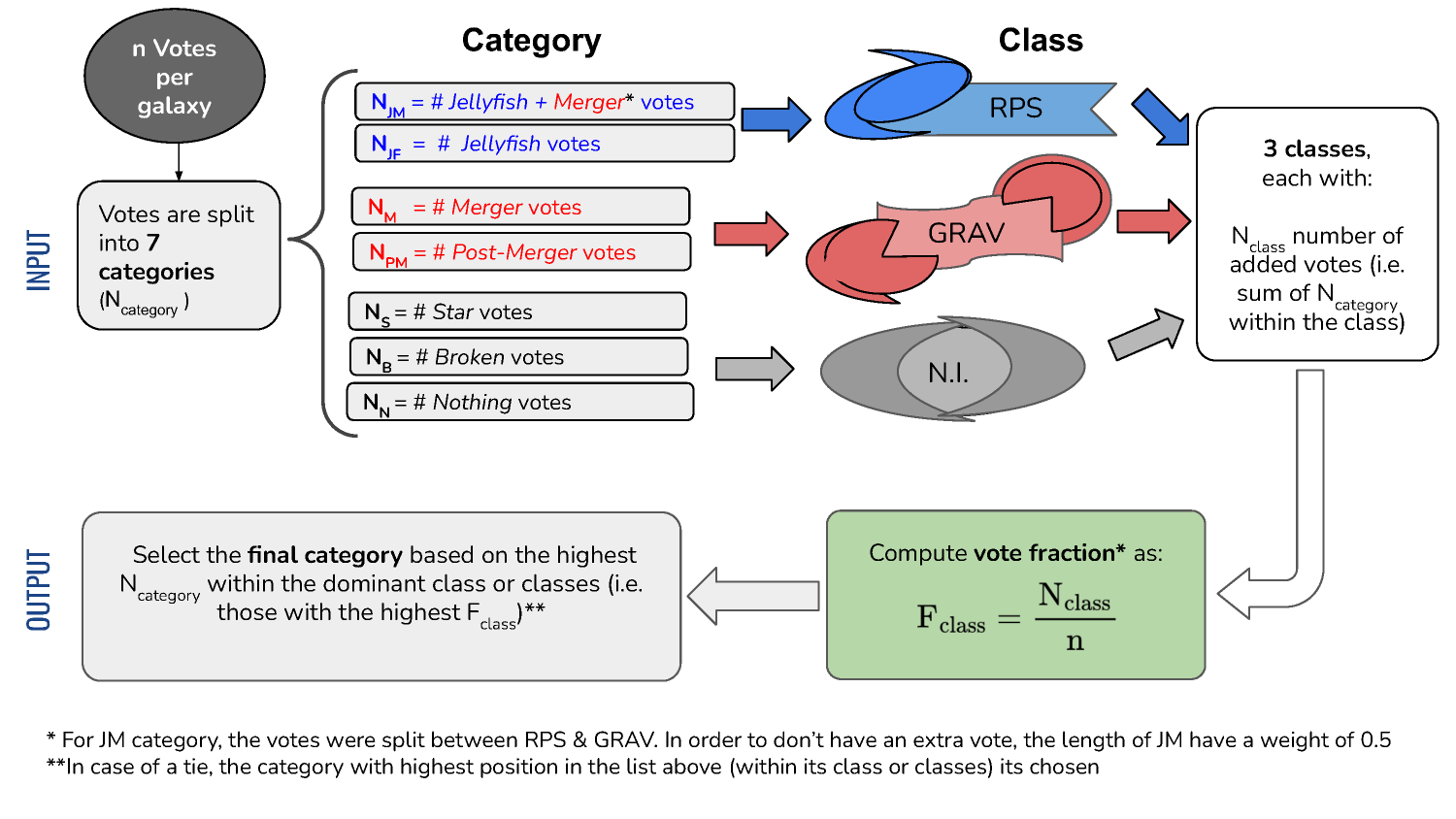}
    \end{center}
    \caption{Flowchart with the visual classification scheme. The \textit{input} consists of the classifications of the different classifiers in different categories, while the \textit{output} corresponds to a unique final category (JF, JM, M, PM, S, B or N), accompanied by the likelihood of being in each of  the 3 main classes (i.e. the vote fractions $F_{\rm RPS}$, $F_{\rm GRAV}$ and $F_{\rm N.I.}$) as explained in the text.
    }
    \label{fig:Fig7}
\end{figure*}

Once all galaxies were classified by at least n = 5 classifiers (where > 22\% of the sample have n = 7), we follow the flowchart in Figure \ref{fig:Fig7} to allocate a final category for each galaxy and associated uncertainty for the type of disturbance (Gravitational, RPS, or none). 

We start by counting the total number of votes per each galaxy ($n$). We then count how many of the $n$ votes went into each of the seven categories ($N_{\rm category}$). Some of the categories are related (e.g. M and PM), so before assigning a final category for the galaxy, we group the votes of related categories into three broader classes. This allows us to compute an uncertainty that reflects the vote fraction for the general class, rather than the specific category. The three classes are:  

\begin{itemize}
    \item \textbf{Ram-pressure stripping class (RPS):} Includes the categories JF and JM. To estimate the likelihood that the galaxy is experiencing RPS, the fraction of the JF and JM votes over the total number of votes is computed as:

    \begin{equation}
        F_{\rm RPS} = \dfrac{N_{\rm JF} + 0.5\times N_{\rm JM}}{n}
        \label{eq:rps}
    \end{equation}

    where $N_{\rm JF}$ is the number of JF votes, and $N_{\rm JM}$ is the number of JM votes.
    
    \item \textbf{Gravitational interactions class (GRAV):} Includes M and PM. Likewise, we compute the likelihood that a galaxy is undergoing a gravitational interaction as the fraction of JM, M, and PM votes, following:
    
    \begin{equation}
        F_{\rm GRAV} = \dfrac{N_{\rm M} + N_{\rm PM} +  0.5\times N_{\rm JM}}{n}
        \label{eq:grav}
    \end{equation}

      where $N_{\rm M}$ is the number of M votes, and $N_{\rm PM}$ is the number of PM votes.
    
    \item  \textbf{Non-Disturbed class (N.I.):}  We finally compute the combined probability that the galaxy is not disturbed, by considering the N, S, and B votes: 

    \begin{equation}
        F_{\rm N.I.} = \dfrac{N_{\rm S} + N_{\rm B} + N_{\rm N}}{n}
        \label{eq:NI}
    \end{equation}

    where $N_{\rm S}$ is the number of S votes, $N_{\rm B}$ is the number of B votes, and $N_{\rm N}$ is the number of N votes.

\end{itemize}

Note that for JM category, the votes were split between the RPS and GRAV class (see Eq. \ref{eq:rps} and \ref{eq:grav}) since in some rare cases the classifiers saw optical disturbances that could be caused by gravitational and/or hydrodynamical effects and it is  not clear which of them dominate. In fact, there are examples in the literature of both effects acting simultaneously on cluster galaxies \citep[e.g. ][]{Moretti2018a}. In these cases, half of the weight was assigned to each vote fraction in order to preserve the total number of votes per galaxy. 

The vote fractions in each class ($F_{\rm class}$) are indicative of  the dominant process affecting (or not) the galaxy and are used as a measure of the uncertainty in the final classification.

\begin{figure*}
    \centering
    \includegraphics[scale=0.2]{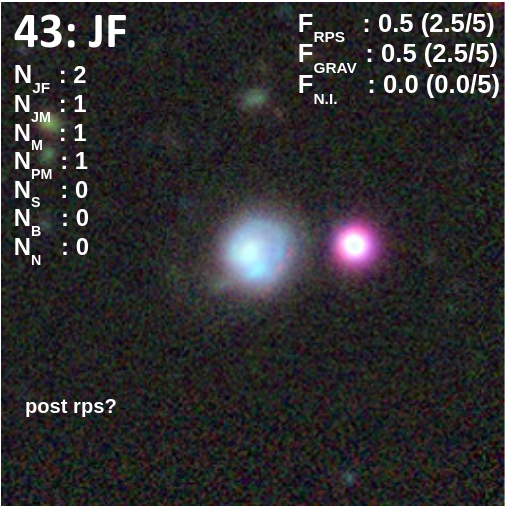}
    \includegraphics[scale=0.2]{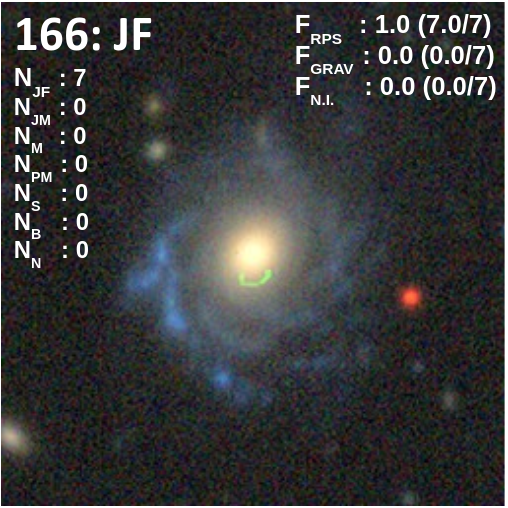}
    \includegraphics[scale=0.2]{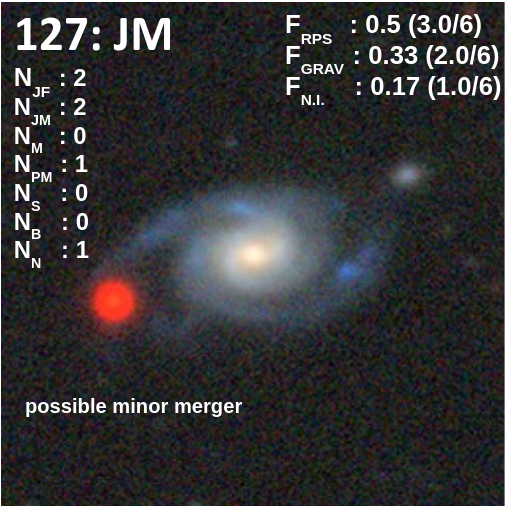}\\
    \includegraphics[scale=0.2]{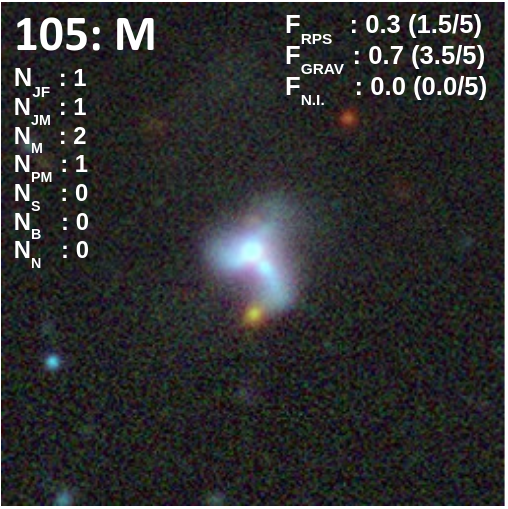}
    \includegraphics[scale=0.2]{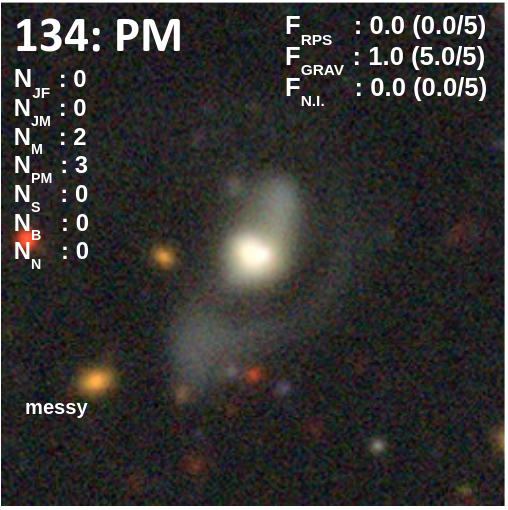}
    \includegraphics[scale=0.2]{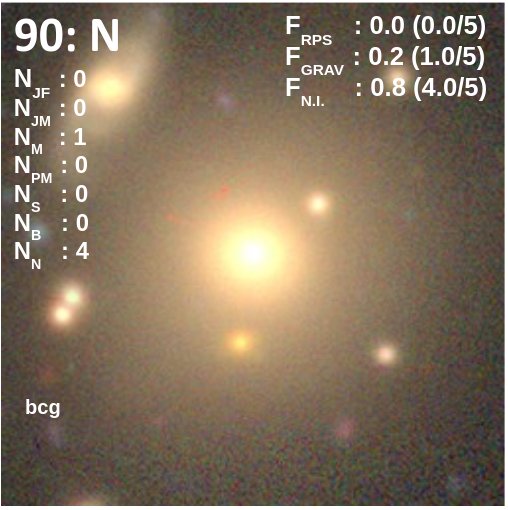}
    \caption{A few examples of different classifications performed to different photometric cluster members. In the upper left, we can see the ID of the galaxy and the final category assigned to each galaxy. Below that, we see the number of votes for each category ($N_{\rm category}$). In the upper right, the vote fraction for each class is overplotted ($F_{\rm class}$). In the bottom left, for some galaxies, we see comments from the classifiers.}
    \label{fig:Fig8}
\end{figure*}

To define the final category of a galaxy, we select the dominant class with the highest vote fraction. If there is one dominant $F_{\rm class}$, the assigned final category will be that with the highest $N_{\rm category}$ within that class. For example, if the vote fraction of the RPS class ($F_{\rm RPS}$) dominates, the category with more votes among JF and JM will be the final category for this galaxy (see examples in Fig. \ref{fig:Fig8}, in the top middle and right panels). For 5\% of photometric cluster members, there was more than one dominant class. In these cases, the final category assigned is that with the highest $N_{\rm category}$ within the dominant classes.

\begin{figure}
    \centering
    \includegraphics[width=0.5\textwidth]{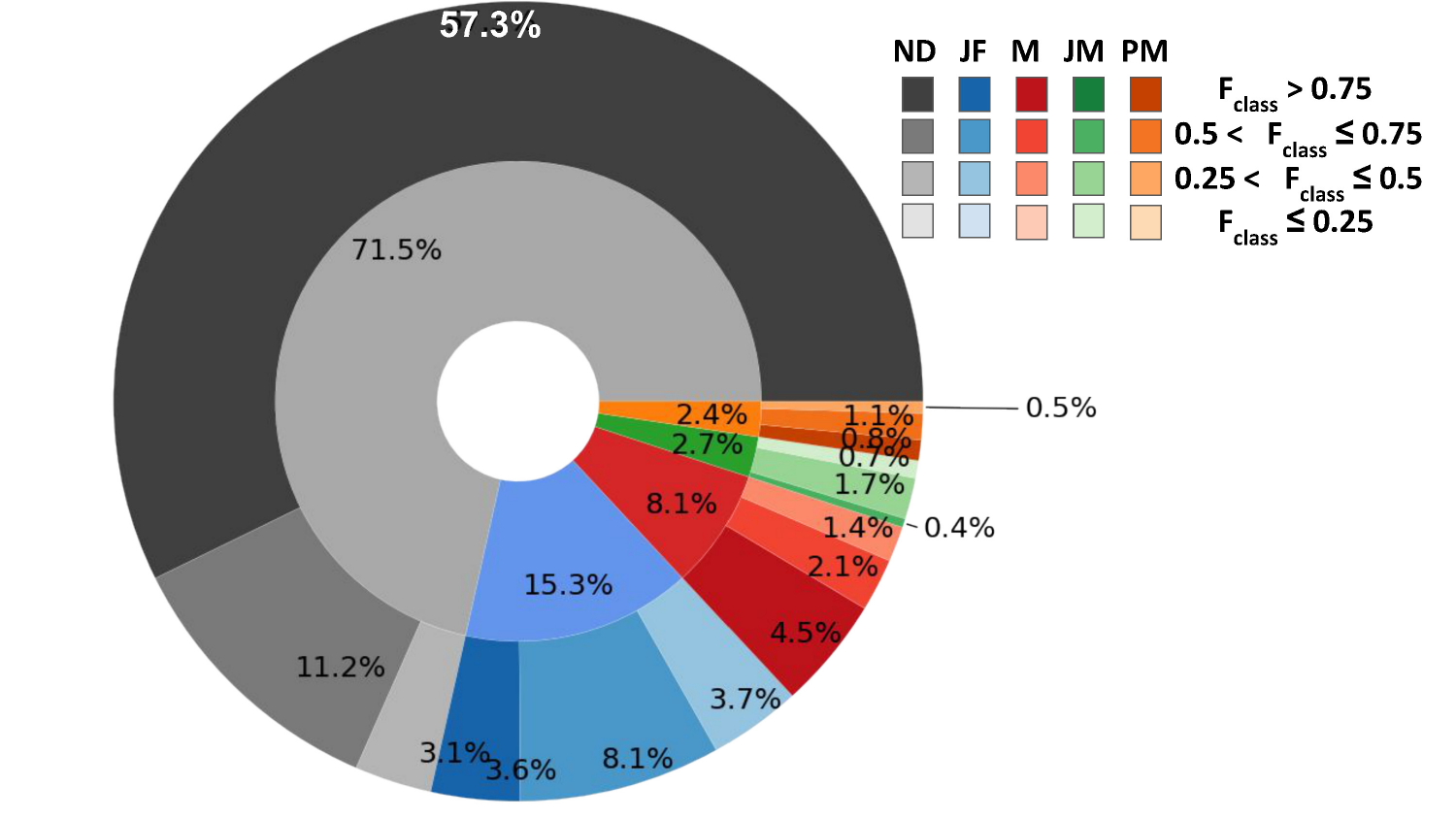}
    \caption{Pie chart of the fraction of disturbed and non-disturbed galaxies in the  photometric cluster sample. \textbf{Inner:} Fraction of galaxies that were classified as JF, M,  JM, PM or ND (Non-disturbed: N, S, B) with respect to the total number of photometric cluster members. \textbf{Outer:}  Fraction of galaxies within a given category that were classified in 4 different ranges of vote fraction values ($F_{class}$) with respect to the total number of photometric cluster members. The distribution  of fractions for the spectroscopic sample (not shown) is very similar.}
    \label{fig:Fig9}
\end{figure}

When assigning the final category, $N_{\rm categories}$ are occasionally tied within the dominant class(es). In general, we gave priority to the categories listed higher in the flowchart of Figure~\ref{fig:Fig7}. For instance, in case of a tie among $N_{\rm JF}$ and $N_{\rm JM}$, we choose the latter as the final category to be conservative (see example in Fig. \ref{fig:Fig8}, top right). If $F_{\rm GRAV}$ dominates, the category with more votes among JM, M, and PM is assigned (Fig. \ref{fig:Fig8}, bottom left and centre). In the case of a tie between $N_{\rm JM}$, $N_{\rm M}$ and/or $J_{\rm PM}$, we assigned the final category JM to ensure we did not throw away possible RPS candidates. If only M and PM share the same amount of maximum votes, we opt for M as the final category. 
If RPS and GRAV have the same (dominant) vote fraction, the category with more votes among JF, JM, M, and PM will be adopted as the final category (Fig. \ref{fig:Fig8}, top left). 
If the vote fraction of N.I. class dominates, the category with more votes among S, B, and N is considered the final one (Fig. \ref{fig:Fig8}, bottom right). In the case of a tie, S was the assigned category. In case B and N have the same amount of maximum votes, B was prioritized.

In summary, the output consists of a final category (JF, JM, M, PM, S, B, or N) for each galaxy, accompanied by the likelihood of that galaxy belonging to each of the three classes (i.e. the vote fractions $F_{\rm RPS}$, $F_{\rm GRAV}$, and $F_{\rm N.I.}$) that indicate the likelihood of a galaxy being disturbed by RPS, gravitational interactions or not being disturbed at all. To visualize the classification scheme, we give some examples of different classifications in Figure \ref{fig:Fig8}. In the top-left example (Galaxy 43), the galaxy has $n=5$ classifications, with two votes for JF, one for JM, one for M, and one for PM. The computation of the RPS vote fraction, $F_{\rm RPS}$, considers only the votes for JF and JM (see eq. \ref{eq:rps}), and yields 0.5. Likewise, $F_{\rm GRAV}$ (Eq. \ref{eq:grav}) considers only the votes for JM, M, and PM, and yields 0.5. Naturally, $F_{\rm N.I.}=0$ (Eq. \ref{eq:NI}) as there were no votes for S, B, or N. Since $F_{\rm RPS} = F_{\rm GRAV}$ for this galaxy, the final classification is that with the most votes within these two classes, which in this case is JF. Therefore the galaxy with ID:43 is a JF candidate with 50\%  likelihood to be hydrodynamically disturbed, 50 \% likelihood to be interacting gravitationally, and 0.0\% likelihood of not undergoing an disturbance or  interaction.

\subsection{The population of \textbf{disturbed} galaxies}
\label{sec:Sec6.2}

In Figure~\ref{fig:Fig9}, the percent of disturbed and non-disturbed galaxies for the entire cluster galaxy sample is illustrated. It is clear that the majority of the sample is not disturbed, but a significant amount (almost 30\%  of all the photometric cluster members)  are. 
Specifically, 240 galaxies are considered to be undergoing gravitational or hydrodynamical disturbances based on their final categories ($=$ JF, JM, M or PM, listed in \ref{tab:tab2}). Being more conservative and only taking into account galaxies with high vote fractions within the disturbances classes (i.e. $F_{\rm RPS}$ or $F_{\rm GRAV} > 0.5$), we find a lower but still impressive number of disturbed galaxies (174), corresponding to $\sim$ 21\% of the photometric cluster members. In addition, 15.3\% (129) and 2.7\% (23) of the photometric cluster members were classified as JF and JM, respectively. That implies that the dominant class with more than half of the disturbed galaxy sample is RPS (over 60\% including JF and JM). However, not all of them have a high vote fraction. This can be seen in the outer pie-chart of Figure~\ref{fig:Fig9}, where it is shown that 11.7\% (98) of photometric cluster member where classified as JF and have a $F_{\rm RPS} > 0.5$, which implies that only over 40\% of disturbed galaxies classified as JF have a high confidence of being real RPS candidates. And if we consider the rare JM cases, the situation naturally worsens, with only 0.4\% (3) of photometric cluster member classified as JM with $F_{\rm RPS} > 0.5$, which implies that petty much none ($\sim 1 \%$) of the JM have a high confidence.

Unless otherwise stated, from now on we will consider disturbed galaxies only those with clear signs of disturbance, i.e. those with disturbed vote fractions (either $F_{\rm RPS}$ or $F_{\rm GRAV}$) higher than 0.5. 
We can thus conclude that $\sim29$\% 
of the cluster galaxies in the photometric cluster member sample are clearly disturbed. 
Of those $\sim 59$\% are being affected by ram-pressure stripping (RPS) and $\sim 41$\%  by gravitational interactions (GRAV). 

\subsubsection{Spatial distribution}
\begin{figure*}          
    \centering
    \includegraphics[width=0.7\textwidth]{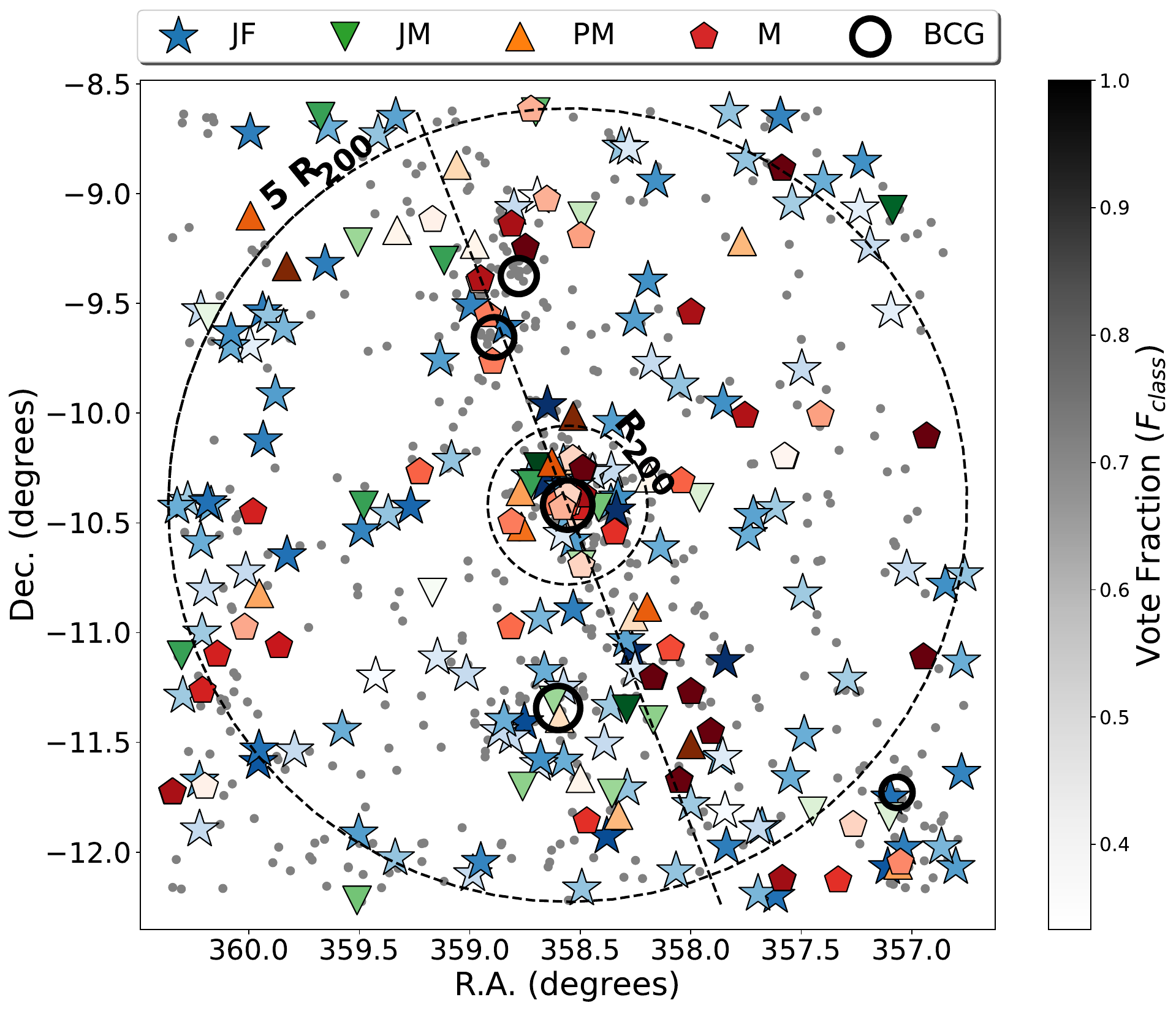}
    \includegraphics[width=0.68\textwidth]{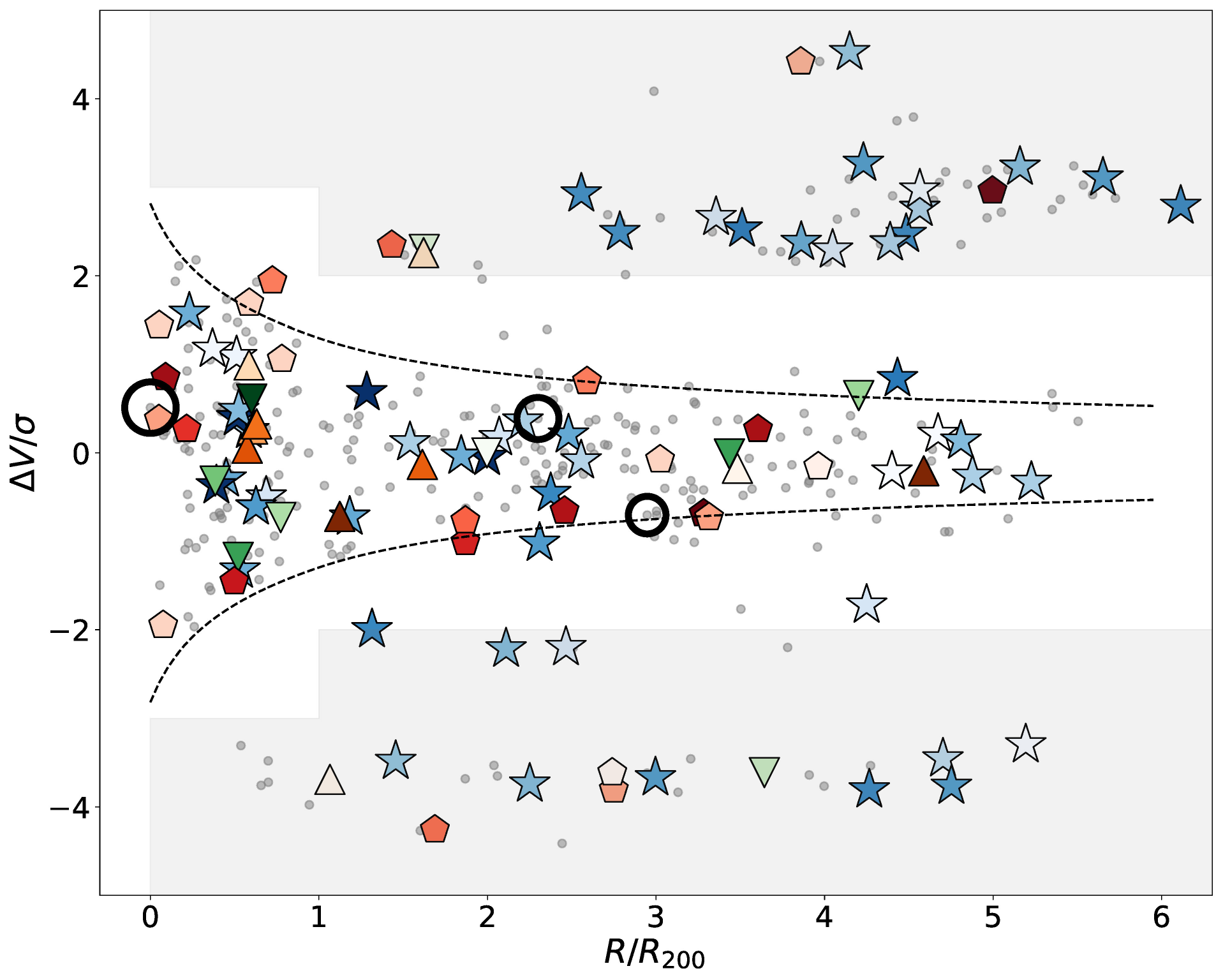}
    \caption{\textbf{Top:} Distribution of all disturbed galaxies identified in this work in the field of A2670. The final classifications (JF, JM, PM, and M) are marked with different symbols and colours. The corresponding vote fractions are indicated by the colour gradient, where the more confident classifications are shown with darker colours, using $F_{\rm RPS}$ for JF and JM galaxies and $F_{\rm GRAV}$ for PM and M galaxies. We use a grey colour bar to represent the gradient of the four colours. The small and large dashed circles indicate $R_{200}$ and $5 \times R_{200}$, respectively. The 5 brightest galaxies are also plotted as larger filled circles, with the sizes reflecting their relative magnitudes (in the phase space we only visualize those with available velocity). The dashed line in the top panel is the same from Fig~\ref{fig:Fig5} to show the merger axis. \textbf{Bottom:} Distribution of all disturbed galaxies with available redshift (corresponding to spectroscopic members) in  position-velocity phase space. The shaded region of unbound galaxies defined in Section~\ref{sec:Sec5.3} is also visualized here. In both panels, the grey points correspond to no disturbed galaxies whose final classification was S, B, or N.} 
    \label{fig:Fig10}
\end{figure*}

The spatial and phase-space distribution of all disturbed galaxies  is shown in Figure~\ref{fig:Fig10}. The final categories (JF, JM, PM, and M) are marked with different symbols and colours. The assigned vote fractions are indicated by the colour gradient, where the more confident classifications are shown with darker colors, using $F_{\rm RPS}$ for JF and JM galaxies and  $F_{\rm GRAV}$ for PM and M galaxies. We use a grey colour bar to represent the gradient of the four colours.

At first glance, the disturbed galaxies seem to follow the dominant substructures identified in Figure~\ref{fig:Fig5} very well, with many of them clustering in the centre, where the main cluster is, and the distributed mostly in the NE-SW direction, where the largest structures (and brightest galaxies) are. 
Focusing on RPS, we find 101 clear candidates (darker blue stars in the Figure~\ref{fig:Fig10}) in the cluster region, 24 of which are spectroscopically confirmed to be bound to the cluster (i.e. excluding the higher velocity galaxies, as described in Sec.~\ref{sec:Sec5.3}). 
A remarkable finding is the presence of clear cases of RPS very far from the cluster core (beyond $2.5 \times R_{200}$, region largely unexplored by previous RPS searches). These are shown in the supplementary material (see Appendix~\ref{ap:exJF}), where there are even some examples of high-confidence RPS candidates located as far as  $> 4\times R_{200}$, including  a few bound cluster galaxies (e.g. Galaxy ID: 842, 843) but mostly unbound e.g. those ones located in the eastern sheet (e.g. ID: 135, 222, 712).

The radial extent of  disturbed galaxies can be better appreciated in Figure~\ref{fig:Fig12}, where the fraction of disturbed galaxies is plotted against the clustercentric distance for bound galaxies only, split by cause of disturbance (RPS vs. GRAV). 
Overall disturbed galaxies are less common with increasing radius.  Interestingly, the fraction of RPS and GRAV galaxies are comparable at all clustercentric distances. 
%

\subsubsection{The effect of substructures}

\begin{figure*}
    \centering
    \includegraphics[width=0.9\textwidth]{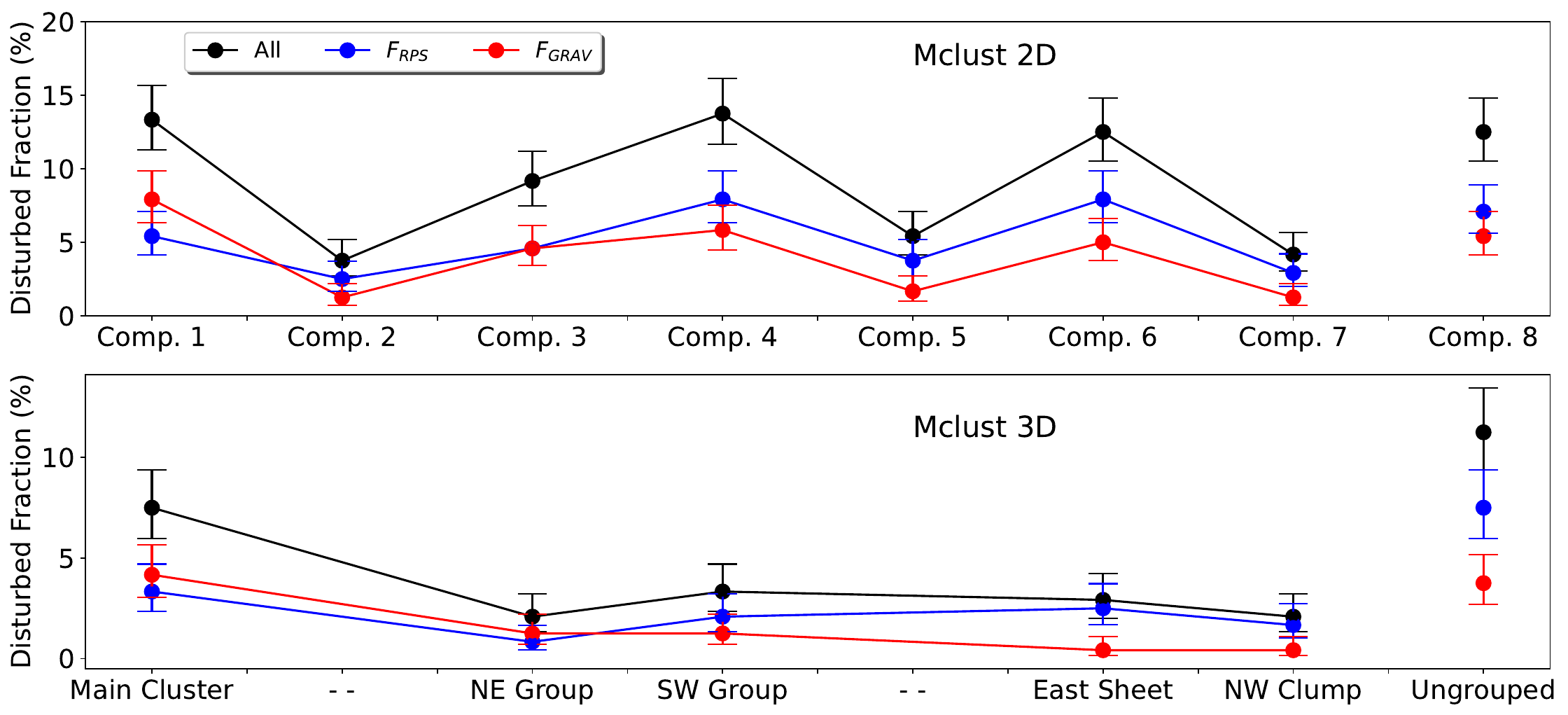}
    \caption{The disturbed galaxy fraction is visualized for each substructure found by \textit{mclust} in 2D (top) and 3D (bottom) \textit{mclust}, ordered from left to right by decreasing density (the ungrouped galaxies are plotted to the right). The black line/symbols correspond to all disturbed galaxies, the blue shows disturbed galaxies  with  $F_{\rm RPS} > 0.5$, and the red those with $F_{\rm GRAV} > 0.5$. The error bars correspond to a binomial proportion confidence interval, using a Wilson interval.}
    \label{fig:Fig11}
\end{figure*}

\begin{figure}
    \centering
    \includegraphics[width=0.48\textwidth]{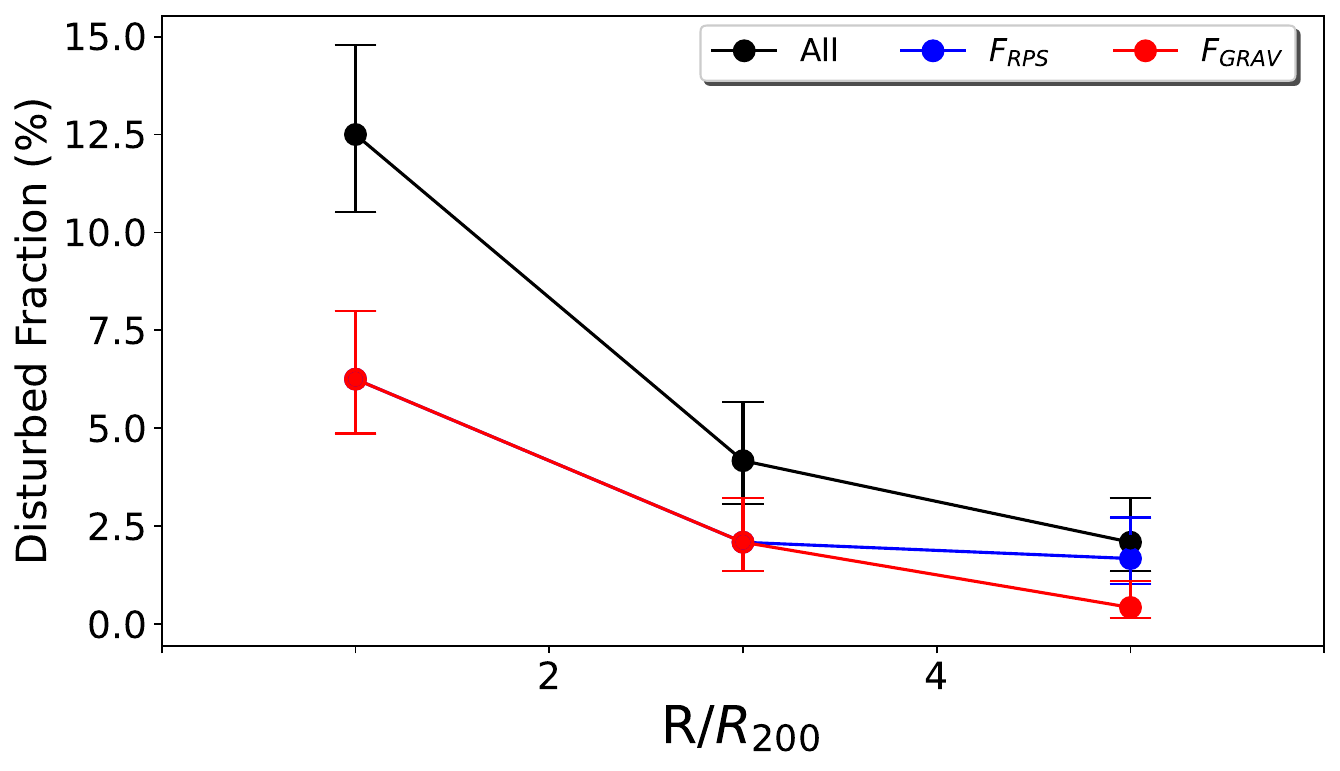}
    \caption{The disturbed galaxy fraction as a function of  clustercentric distance for bound cluster galaxies only (defined in Sec.~\ref{sec:Sec5.3}). 
    The  solid line/symbols corresponds to all the disturbed galaxies (black),  RPS candidates (blue)  Gravitational interacting candidates (red). As in other plots we only consider disturbed galaxies with vote fractions above 0.5. 
    The error bars correspond to a binomial proportion confidence interval, using a Wilson interval.}
    \label{fig:Fig12}
\end{figure}

To investigate the correlation between galaxy morphology and cluster substructures, in Figure \ref{fig:Fig11} we plot the fraction of disturbed galaxies (black line) as a function of cluster substructures, ordered by decreasing density. The top panel corresponds to structures identified by \textit{mclust} 2D and the bottom panel to \textit{mclust} 3D. In the 3D case, where substructures are more robustly identified, we find that the main cluster has the highest number of disturbed galaxies, and the substructures that follow in density have increasingly lower fractions of disturbed galaxies. The case of 2D is not too dissimilar if we exclude substructures that are very loose, poorly defined, or at the edge of the field where the spectroscopic information is not available (namely Comp.~2, 4, and 5). 
Interestingly, in both cases the fraction of disturbed galaxies outside of substructures (in Comp.~8 or in the "Ungrouped" component) is comparable or even higher than that in the main cluster, but note that many of these are not really associated (bound) to the cluster. 

We split the disturbed galaxies into hydrodynamical and gravitational interactions by considering $F_{\rm RPS} > 0.5$  and $F_{\rm GRAV} > 0.5$, respectively (blue and red lines in Figure~\ref{fig:Fig11}). We find that gravitational interactions (80 \% mergers and 20\% post mergers) are comparable with the RPS effects in the cluster core within errors. 
Interestingly, in lower density (unbound) structures towards the cluster outskirts (at $5\times R_{200}$), the fraction of galaxies experiencing RPS is $\sim 4$ times larger than that of gravitationally interacting galaxies (over 1$\sigma$ significance). 

However, if we explicitly exclude  galaxies that are not bound to the cluster (as defined in Sec.~\ref{sec:Sec5.3}) we find that the fraction of disturbed galaxies in the Main Cluster (10.8\%) is slightly higher than that inside of substructures (8.8\% in the spectroscopic sample) and both are at the same time comparable to the fraction outside of substructures (6.7\%) within uncertainties\footnote{Note that these fractions are not directly relatable to Fig.\ref{fig:Fig11} as they exclude unbound galaxies while the Figure considers both bound and unbound}. This result highlights the complexity of galaxy evolution in assembling clusters (see discussion in Sec.~\ref{sec:disc}).

\subsubsection{The effect of the cluster merger}

\begin{figure}
    \includegraphics[width=0.48\textwidth]{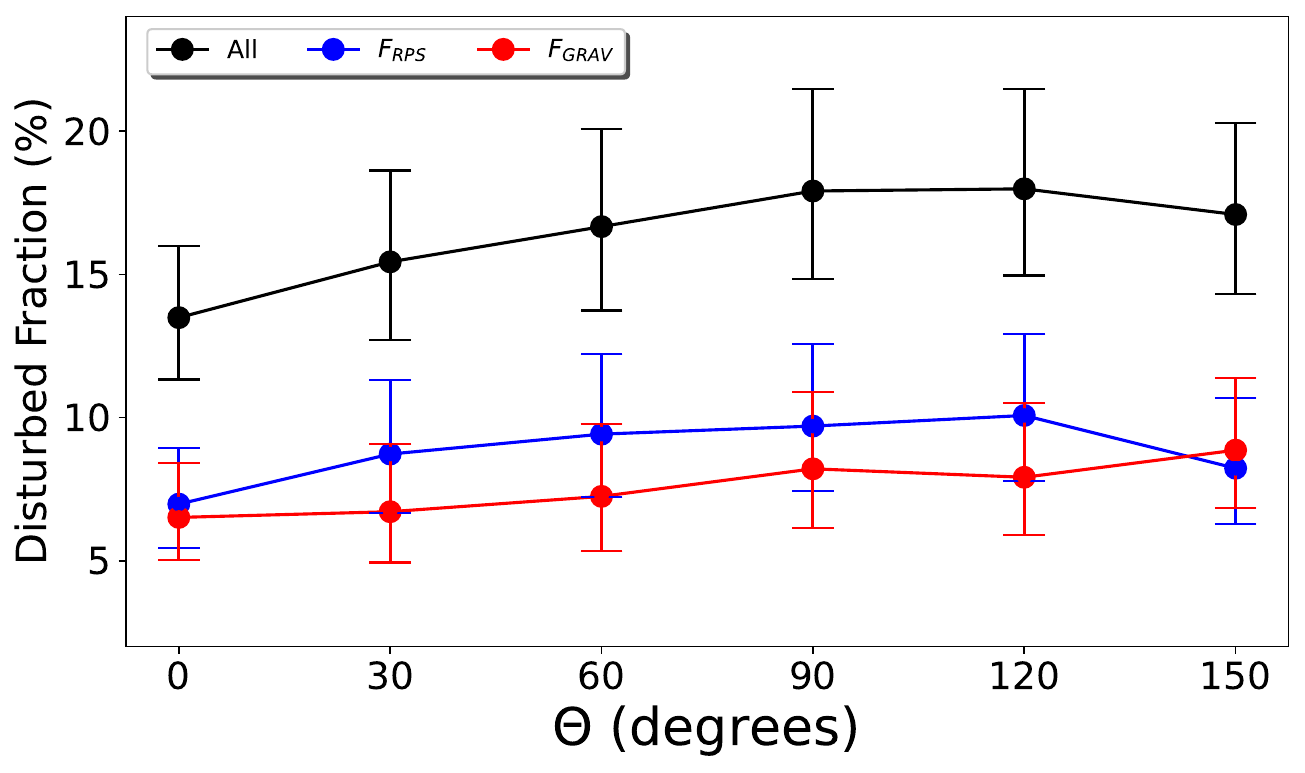}
    \caption{The disturbed galaxy fraction in rectangular regions centered on the BCG and rotated relative to the cluster merger axis. The angle $\theta$ is zero when the rectangle is aligned with the cluster merger axis, and $90\deg$ when its perpendicular to it. 
    As in Figure~\ref{fig:Fig12}, only galaxies bound to the cluster were considered. All disturbed galaxies are shown in black, RPS candidates in blue and gravitational interactions in red. Again, only disturbed galaxies with high vote fractions were considered. The error bars correspond to a binomial proportion confidence interval, using a Wilson interval.}
    \label{fig:merg_angle}
\end{figure}

Finally, in an attempt to test the role of post-processing more directly, we quantified the number of disturbed galaxies along the merger axis which connects the NW Group with the Main cluster (see dashed line in Figure~\ref{fig:Fig9}) in a rectangular region ($0.7^{\circ}$~wide) centered on the BCG and aligned with the cluster merger axis. We found that compared to any other rotation this axis has the highest number of disturbed galaxies. In particular, we find that out of all the clear cases of gravitationally interacting galaxies (M, PM) in the  photometric sample of A2670, 16\% are in this direction. Likewise, from the the potential RPS candidates (JF and JM), 10\% are located along the cluster merger axis. The large number of disturbed galaxies in this axis is however a consequence of the higher galaxy density in this region. In fact, when we compute the  fraction of disturbed galaxies over all the galaxies along the merging axis and 5 other rotated regions (shown in Figure~\ref{fig:merg_angle}), we find no significant difference. If anything we find a slight increase in the fraction of disturbed galaxies (mostly RPS candidates) in the direction perpendicular to the merger axis, but with very low significance. 

At face value the lack of significant correlation between the disturbed galaxy fraction and the location relative to the cluster merger axis seems to indicate, that the cluster merger is not accelerating galaxy evolution in any particular direction.

\section{Discussion}
\label{sec:disc}

\subsection{Pre- and post-processing:}
In Section~\ref{sec:Sec6} we found a remarkably high number (174) of  
disturbed galaxies in A2670 and its surroundings. The majority of those 
are likely being disturbed by ram-pressure stripping, while the rest are likely disturbed gravitationally. 
Our study reports the highest number of RPS candidates to date in a single system ($101$). Prior to this study, the record was held by \citep[][]{RomanOliveira2019, Ruggiero2019} who claim to have found 70 RPS  candidates in a similar system: A901/2, composed of two galaxy clusters and two smaller groups in route of collision with each other simultaneously. 
These two works combined could suggest that dynamically young clusters can indeed enhance RPS \citep[see also][]{Owers2012}. These single-system studies seem to be at odds with a recent study \citep[][]{Lourenco2023} aiming at quantifying the effect of cluster dynamical state on RPS. Using the large homogeneous sample of RPS candidates from \citep{Poggianti2016}, they find no correlation between the fraction of RPS candidates relative to the infalling population of blue spirals and cluster dynamical stage. However, their results could be affected by the limited area covered by the observations ($0.7 \times R_{200}$). Our findings thus motivate larger-sample studies such as that of \citep[][]{Lourenco2023} but extending well beyond the virial radius of clusters.


To understand how unique A2670 is, we compare our results with statistical studies of the incidence of optically-selected RPS candidates in clusters. The most recent one is \citep[][]{Vulcani2022}, who found that these galaxies represent $\sim15-20\%$ of the infalling population of blue spiral galaxies (within roughly the virial region) in a large sample of mostly relaxed clusters from the WINGS/OmegaWINGS surveys \citep[similar to][]{Lourenco2023}. In order to compare in a meaningful way our results with this study we consider only the blue galaxies in A2670 (those below 1$\sigma$ from the red sequence line in Fig. \ref{fig:FigA2}) and assume they are mostly spirals. We find that between $\sim22 - 32 \%$ of the blue members of A2670 (up to $5\times R_{200}$) are being affected by RPS, which is indeed higher than what is found by \citet[][]{Vulcani2022}. The fraction of blue RPS candidates in A2670 decreases only very mildly (21-30\%) when we restrict the sample to $1\times R_{200}$ (closer to the area explored by \citet[][]{Vulcani2022}). 

When inspecting the distribution of the (bound) disturbed galaxies in A2670 we find that, for both gravitational or RPS cases, the fraction of disturbed galaxies decreases with increasing clustercentric distance. 
In the case of RPS candidates the results is consistent with hydrodynamical simulations which suggest gas stripping can start to take place even at $\sim 5 \times R_{200}$ \citep{Bahe2013} and that the peak of RPS is located around the virial radius \citet{Pallero2022}.


The increasing fraction of gravitationally interacting galaxies towards the cluster center is harder to explain given that galaxy mergers are not as likely to happen in cluster cores due to the higher relative motions of galaxies . In merging systems however asymmetric galaxies have been found to be more common near the core \citep[e.g.][]{Kleiner2014}, and indeed the distribution of merging and particularly post-merger galaxies in A2670 is concentrated along the merger axis where the galaxy density is generally higher.  
We interpret these results as an effect of post-processing at play, where the dynamically active environment and in particular the merger of the different structures enhances galaxy-galaxy interactions (as well as RPS) in unexpected regions.

The general finding is that almost 40\% of the disturbed galaxies are located well beyond the virial region at $> 4 \times R_{200}$ (in all directions). Most of these are RPS candidates in substructures,
implying that RPS could be an important mechanism transforming galaxies even in groups outside the cluster (pre-processing). 
\citet[][]{Pallero2019,Pallero2022} indeed show that the fraction of preprocessed galaxies is expected to be larger for massive clusters like A2670. 
In addition, A2670 is in the process of growth, with several well defined structures falling into it, which could already be processing galaxies if the structures are massive enough \citep{Pallero2022}. 
We further investigate the local environment of galaxies to try to catch pre-processing at play: We naively consider the disturbed galaxies in groups to be a consequence of pre-processing and those in between the structures (especially along the merging axis) as post-processing. Although most disturbed galaxies are in substructures along the merging axis, the relative abundance is similar in and out of substructures, which indicates galaxies are likely being affected by the cluster, the groups and possibly the accretion process of the cluster, highlighting the complex nature of this system.

In summary, our study showcases the importance of pre- and post-processing of galaxies in actively growing clusters and presents a motivation for larger homogeneous studies of galaxies within their large scale-environment.


\subsection{Caveats and future improvements}

One of the main caveats of our study is the use of broad-band optical images, which are not as sensitive as e.g. radio (HI) or H$\alpha$ imaging for detecting RPS features. However, they are more widely available which allows to search for these kinds of features in large areas of the sky. And although optical images only see a small part of the RPS process, the GASP survey \citep{Poggianti2017a} has shown based on MUSE data for $>$100 optically-selected RPS candidates, 
that the "success rate" in finding real RPS galaxies using optical images in clusters is 86\% (Poggianti et al. in prep). We can assume that in our sample the real cases of RPS are of the same order. However, there is a possibility that outside the cluster core the contamination from non-RPS galaxies is higher (GASP does not probe beyond $\sim 2\times R_{200}$). To show that the RPS signatures we find in the cluster outskirts are not (visually) different from those found in the central regions of clusters \citep[by this work and also works like][]{Poggianti2016}, 
we show all the galaxies classified as JF with high vote fraction in the cluster up to $5 \times R_{200}$ in the supplementary material described in \ref{ap:exJF}. We also note that the vote fraction for any given category is constant across clustercentric distances.  

Another likely source of bias in our study is the possible dependence of our results with galaxy mass. Due to low number statistics we do not bin in stellar mass, but we note that the mass distribution of the galaxies in the different structures is quite similar.

Finally, while our study is wide, it only considers a single system. Future studies covering a wide regions around multiple clusters in multiple wavelengths will be able to increase the statistics and better probe the different mechanisms transforming galaxies in the different local environments. 
The measurement of tail directions in RPS candidates (see e.g. Salinas et al. submitted) can also help determine and separate the effects of pre- and post-processing in merging/disturbed environments and we plan to study this in a future study.

\section{Summary and conclusions}
\label{sec:Sec7} 

In order to understand the complex role of the evolving environment in galaxy transformations (in particular the role of \textit{pre} and \textit{post-processing}), we studied the properties of galaxies in a large area centred on the interacting system A2670, a nearby massive cluster of galaxies. We looked for signs of transformations in galaxies not only in the inner part of the cluster where many galaxy evolution studies focus, but also in the infall regions of the cluster up to $\sim$ 5$\times R_{200}$ where galaxies are falling in from the cosmic web. 

We first built a complete and homogeneous spectro-photometric catalogue using public data from LS DR9 and the SDSS when available for a $3.6^{\circ} \times  3.6^{\circ}$ region centred on A2670. We select photometric and spectroscopic members by considering galaxy colours and redshifts respectively within a magnitude limit of $M_B < -20$. 

We then studied the distribution and clustering of galaxies in this large region containing the cluster and its infall regions using different 2D and 3D techniques, including a Dressler Schectman's test and a gaussian mixture model technique (\textit{mclust}) in 2D and 3D. We could robustly confirm  that the cluster is interacting and has a significant amount of substructure, not only in the core \citep[as found by][]{Hobbs1997,Fujita2006,Sheen2012} but out to the outskirts. 
The substructure analysis clearly identified the main cluster at the centre of the field, accompanied by a large group towards the NE, a group towards the SE, 2 lower-density high-velocity substructures far from the cluster centre, and other less-defined structures. Our results are consistent with similar but smaller-area studies of this cluster  \citep[e.g.][]{LopezAlfaro2022}, that have found at least three groups running along the NE-SW axis with a high probability to be merging with the main cluster.  Further confirmation comes from the 5 brightest galaxies in the system, which are aligned with the merging axis.  

After characterizing the environment in this cluster, we visually inspected the member galaxies in search of signatures of ongoing or past gravitational (mergers and post-mergers) and hydrodynamical disturbances (RPS). We used optical images from LS as well as our own deeper DECam data. 
We developed our own interactive tool to allow different users to classify the galaxies, and a method to combine the votes into a final category with associated uncertainty. 
Our tools and methods can serve as a reference for ongoing or future larger studies aiming at identifying disturbed galaxies in large optical datasets visually (e.g. the ongoing citizen science project "\textit{Fishing for Jellyfish Galaxies}"\footnote{\url{https://www.zooniverse.org/projects/cbellhouse/fishing-for-jellyfish-galaxies}}.

A total of 174 clearly disturbed galaxies were found in A2670 and its surroundings. Of those, most ($101$) are ram-pressure stripping candidates, while the rest are likely disturbed gravitationally. 
Our study reports the highest number of RPS candidates to date in a single system, supporting previous claims that cluster mergers could enhance ram-pressure, and trigger galaxy interactions. 

When studying the distribution of the disturbed galaxies in A2670 we find that while many of them agglomerate near the inner part of the cluster, most of them are located  in the outer parts, in and around the substructures, with a high concentration in the merging axis of the A2670 system. 
In particular, almost 40\% of the disturbed galaxies (mostly RPS candidates) are located well beyond the virial region at $> 4 \times R_{200}$, many of which are in substructures (not all bound to the cluster). 
This result highlights the important role that pre-processing must have to transform galaxies prior to entering the cluster. 
If we limit the sample to those bound to the cluster we find that the fraction of disturbed galaxies decreases by a factor of $\sim 3$ from the core to the outskirts of the cluster, where its non-zero. This finding supports the (theoretical) notion that gas removal in particular can start very far from the virial region of the cluster \citep{Bahe2013}, and highlights the transformational power of clusters even at  large distances.

We also find that the fraction of disturbed galaxies in- and out-side substructures is similar and also comparable to that inside the main cluster. We interpret this finding as a possible combination of pre-processing with post-processing induced by the cluster merger. 
More direct support for the post-processing scenario is the finding that gravitationally interacting galaxies being are common close to the cluster core, and this is rare in regular clusters. 
Although generally there are more disturbed galaxies along the cluster merging axis, when  we compute the fraction of disturbed galaxies along this axis, we do not find any enhancement relative to any other orientation, which disfavours a scenario of accelerated galaxy evolution in a preferred spatial direction for this system.

Overall, our findings show clear environmental influence in galaxies out to 5 virial radii and reflect the complex interplay between the different environmental effects in the dynamically young cluster of A2670. On one hand pre-processing in galaxy groups can start to transform galaxies prior entering the cluster, and on the other hand the merging cluster environment can trigger hydrodynamical disturbances and gravitational interactions,  accelerating galaxy evolution (post-processing). Finally, our results also highlight the importance of studying the outskirts of clusters and motivate larger homogeneous studies of galaxies within their large scale-environment. 

\vspace{0.2cm}

\section*{Data Availability}
The data used in this manuscript are primarily sourced from publicly available data from  SDSS DR9 \citep[][]{SDSS_DR9}, Legacy Survey \citep[LS;][]{Dey2019} public data release (DR9), and proprietary deeper DECam imaging (Kim et al. submitted). 
The python tool developed to inspect and classify each galaxy is available at \url{https://github.com/FrancoPiraino/Interacting_Galaxy_Classification.git}.
All the classifications performed for disturbed galaxies are available in Table~\ref{tab:tab2}.

\section*{Acknowledgements}
We acknowledge support from Eduardo Ibar and FONDECYT Regular project No. 1221846, and thank Patricia Ar\'evalo and Sergio Torres-Flores for their useful comments on the manuscript.
FPC acknowledges financial support from Direcci\'on de Postgrado (Universidad T\'ecnica Federico Santa Mar\'ia, Chile) through Becas Internas para Doctorado y Mag\'ister Cient\'ifico-Tecnol\'ogicos.
FPC also acknowledges partial financial support from Centro de Astrof\'isica de Valpara\'iso - CAV, CIDI No. 17 (Universidad de Valpara\'iso, Chile), the Postgrado en Astrof\'isica program (Universidad de Valpara\'iso, Chile) and the financial support from FONDECYT Iniciaci\'on No. 11180558.
YLJ, ACL and DP acknowledge financial support from ANID BASAL project No. FB210003. 
YLJ also acknowledges FONDECYT Regular No. 1230441.
ACL thanks the financial support of the National Agency for Research and Development (ANID) / Scholarship Program / DOCTORADO BECAS CHILE/2019-21190049.
JPC acknowledges financial support from ANID through FONDECYT Postdoctorado Project 3210709.
DK acknowledges support from the Korea Astronomy and Space Science Institute under the R\&D
program (Project No. 2022-1-868-04) supervised by the Ministry of Science and ICT, and the National Research Foundation of Korea (NRF) grant funded by the Korean government (MSIT) (No. NRF-2022R1C1C2004506).
YKS acknowledges support from the National Research Foundation of Korea (NRF) grant funded by the Ministry of Science and ICT (NRF-2019R1C1C1010279). KK acknowledges full financial support from ANID through FONDECYT Postdoctorado Project 3200139. DP acknowledges financial support from ANID through FONDECYT Postdoctorado Project 3230379.
This research made use of the "K-corrections calculator" service available at http://kcor.sai.msu.ru/
Funding for SDSS-III has been provided by the Alfred P. Sloan Foundation, the Participating Institutions, the National Science Foundation, and the U.S. Department of Energy Office of Science. The SDSS-III website is \url{http://www.sdss3.org/.}

SDSS-III is managed by the Astrophysical Research Consortium for the Participating Institutions of the SDSS-III Collaboration including the University of Arizona, the Brazilian Participation Group, Brookhaven National Laboratory, Carnegie Mellon University, University of Florida, the French Participation Group, the German Participation Group, Harvard University, the Instituto de Astrofisica de Canarias, the Michigan State/Notre Dame/JINA Participation Group, Johns Hopkins University, Lawrence Berkeley National Laboratory, Max Planck Institute for Astrophysics, Max Planck Institute for Extraterrestrial Physics, New Mexico State University, New York University, Ohio State University, Pennsylvania State University, University of Portsmouth, Princeton University, the Spanish Participation Group, University of Tokyo, University of Utah, Vanderbilt University, University of Virginia, University of Washington, and Yale University. 

The Legacy Surveys consist of three individual and complementary projects: the Dark Energy Camera Legacy Survey (DECaLS; Proposal ID \#2014B-0404; PIs: David Schlegel and Arjun Dey), the Beijing-Arizona Sky Survey (BASS; NOAO Prop. ID \#2015A-0801; PIs: Zhou Xu and Xiaohui Fan), and the Mayall z-band Legacy Survey (MzLS; Prop. ID \#2016A-0453; PI: Arjun Dey). DECaLS, BASS and MzLS together include data obtained, respectively, at the Blanco telescope, Cerro Tololo Inter-American Observatory, NSF's NOIRLab; the Bok telescope, Steward Observatory, University of Arizona; and the Mayall telescope, Kitt Peak National Observatory, NOIRLab. Pipeline processing and analyses of the data were supported by NOIRLab and the Lawrence Berkeley National Laboratory (LBNL). The Legacy Surveys project is honored to be permitted to conduct astronomical research on Iolkam Du'ag (Kitt Peak), a mountain with particular significance to the Tohono O'odham Nation.
NOIRLab is operated by the Association of Universities for Research in Astronomy (AURA) under a cooperative agreement with the National Science Foundation. LBNL is managed by the Regents of the University of California under contract to the U.S. Department of Energy.
This project used data obtained with the Dark Energy Camera (DECam), which was constructed by the Dark Energy Survey (DES) collaboration. Funding for the DES Projects has been provided by the U.S. Department of Energy, the U.S. National Science Foundation, the Ministry of Science and Education of Spain, the Science and Technology Facilities Council of the United Kingdom, the Higher Education Funding Council for England, the National Center for Supercomputing Applications at the University of Illinois at Urbana-Champaign, the Kavli Institute of Cosmological Physics at the University of Chicago, Center for Cosmology and Astro-Particle Physics at the Ohio State University, the Mitchell Institute for Fundamental Physics and Astronomy at Texas A\&M University, Financiadora de Estudos e Projetos, Fundacao Carlos Chagas Filho de Amparo, Financiadora de Estudos e Projetos, Fundacao Carlos Chagas Filho de Amparo a Pesquisa do Estado do Rio de Janeiro, Conselho Nacional de Desenvolvimento Cientifico e Tecnologico and the Ministerio da Ciencia, Tecnologia e Inovacao, the Deutsche Forschungsgemeinschaft and the Collaborating Institutions in the Dark Energy Survey. The Collaborating Institutions are Argonne National Laboratory, the University of California at Santa Cruz, the University of Cambridge, Centro de Investigaciones Energeticas, Medioambientales y Tecnologicas-Madrid, the University of Chicago, University College London, the DES-Brazil Consortium, the University of Edinburgh, the Eidgenossische Technische Hochschule (ETH) Zurich, Fermi National Accelerator Laboratory, the University of Illinois at Urbana-Champaign, the Institut de Ciencies de l'Espai (IEEC/CSIC), the Institut de Fisica d'Altes Energies, Lawrence Berkeley National Laboratory, the Ludwig Maximilians Universitat Munchen and the associated Excellence Cluster Universe, the University of Michigan, NSF's NOIRLab, the University of Nottingham, the Ohio State University, the University of Pennsylvania, the University of Portsmouth, SLAC National Accelerator Laboratory, Stanford University, the University of Sussex, and Texas A\&M University.
BASS is a key project of the Telescope Access Program (TAP), which has been funded by the National Astronomical Observatories of China, the Chinese Academy of Sciences (the Strategic Priority Research Program "The Emergence of Cosmological Structures" Grant \# XDB09000000), and the Special Fund for Astronomy from the Ministry of Finance. The BASS is also supported by the External Cooperation Program of Chinese Academy of Sciences (Grant \# 114A11KYSB20160057), and Chinese National Natural Science Foundation (Grant \# 12120101003, \# 11433005).
The Legacy Survey team makes use of data products from the Near-Earth Object Wide-field Infrared Survey Explorer (NEOWISE), which is a project of the Jet Propulsion Laboratory/California Institute of Technology. NEOWISE is funded by the National Aeronautics and Space Administration.
The Legacy Surveys imaging of the DESI footprint is supported by the Director, Office of Science, Office of High Energy Physics of the U.S. Department of Energy under Contract No. DE-AC02-05CH1123, by the National Energy Research Scientific Computing Center, a DOE Office of Science User Facility under the same contract; and by the U.S. National Science Foundation, Division of Astronomical Sciences under Contract No. AST-0950945 to NOAO.


\bibliography{ms}
\bibliographystyle{mnras}



\appendix

\section{Galaxies classified as JF in the inner parts and outer parts of the cluster}
\label{ap:exJF}

In the online supplementary material we include 2 figures showing the JF galaxies with high RPS vote fractions located inside the cluster ($r<2.5\times R_{200}$) and beyond (between $2.5$ and $5\times R_{200}$), sorted from high to low $F_{RPS}$ values in each case. The individual figures correspond to $60\arcsec \times 60\arcsec$ RGB image cutouts from LS sky  browser\footnote{\url{https://www.legacysurvey.org/viewer}} \citep[][]{Dey2019} public data release (DR9).

\section{Defining blue and red galaxies}

\begin{figure}
    \centering
    \includegraphics[width=0.48\textwidth]{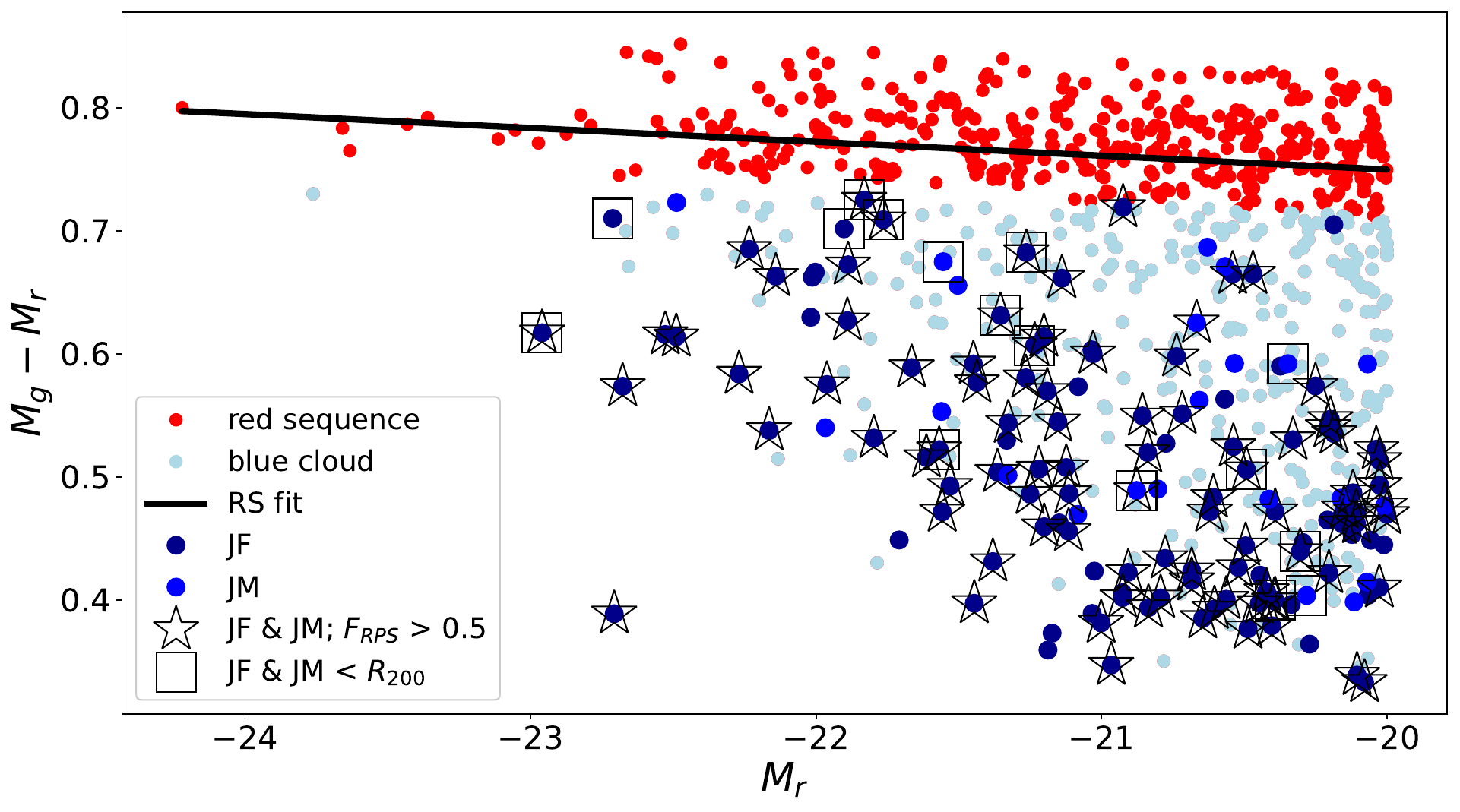}
    \caption{Color magnitude diagram for the photometric cluster members. 
    The red sequence is shown by the black line. 
    The separation between \textit{red sequence} (red dots) and the \textit{blue cloud} (blue markers) galaxies is defined by the 1$\sigma$ line below red sequence as in \citet[][]{Vulcani2022}. The \textit{blue cloud} is composed both by galaxies suffering gravitational interactions or non-disturbed (light blue dots), galaxies that were classified as JF (dark blue dots), and galaxies that were classified as JM (blue dots). The blue galaxies classified both as JF and JM  with vote fraction $F_{\rm RPS} > 0.5$ are plotted as stars and those that are located to < $R_{200}$ are plotted as squares. Our results show that between $\sim 22 - 32$\% of the blue members of A2670 are being affected by RPS and if we restrict the sample to $1\times R_{200}$  \citep[closer to the area explored by][]{Vulcani2022} the fraction of blue RPS candidates decrease only very mildly ($21 - 30$\%).}  
    \label{fig:FigA2}
\end{figure}

In this appendix, we can visualize the colour-magnitude diagram for the photometric cluster members (Figure \ref{fig:FigA2}). The colour bimodality is defined by the separation between the red sequence and the blue cloud, which is composed of all of these galaxies located $1\sigma$ below the red sequence line (black line), yielding a total of 441 blue galaxies. If we restrict the sample to $1\times R_{200}$, it yields a total of 53 blue galaxies.

\section{Final classification for all disturbed galaxies}

\onecolumn
\begin{longtable}{lllllll}
\caption{The fist 10 rows of the table with the final classification for disturbed galaxies. Columns are: ID of the galaxies in our catalogue, coordinates of the galaxies in degrees from LS \citep[][]{Dey2019} public data release (DR9), final category assigned to the galaxy after combining the different votes, vote fraction for each class, $F_{\rm RPS}$, $F_{\rm GRAV}$ and $F_{\rm N.I.}$. The complete version of this table is available in the online version of the paper.}
\label{tab:tab2}
\\ \hline
ID & R.A. & Dec. & Final Category & $F_{\rm RPS}$ & $F_{\rm GRAV}$ & $F_{\rm N.I.}$ \\ \hline
\endfirsthead
\multicolumn{7}{c}%
{{\bfseries Table \thetable\ continued from previous page}} \\
\endhead
7 & 358.49480 & -10.69413 &          M &   0.100 &    0.500 &  0.400 \\
9 & 358.53481 & -10.57998 &          JF &   0.714 &    0.143 &  0.143 \\
13 & 358.49525 & -10.69191 &          JM &   0.300 &    0.700 &  0.000 \\
28 & 358.72571 & -10.28808 &          JF &   0.600 &    0.000 &  0.400 \\
30 & 358.69096 & -10.24600 &          JM &   0.700 &    0.300 &  0.000 \\
33 & 359.08307 & -10.21275 &          JF &   0.600 &    0.000 &  0.400 \\
34 & 358.26567 & -11.08203 &          JF &   1.000 &    0.000 &  0.000 \\
35 & 357.02268 & -10.71181 &          JF &   0.500 &    0.000 &  0.500 \\
40 & 358.04252 & -10.30951 &           M &   0.000 &    0.714 &  0.286 \\
43 & 359.01519 & -11.18858 &          JF &   0.500 &    0.500 &  0.000 \\
\end{longtable}

\bsp	
\label{lastpage}

\ifarXiv
    \foreach \x in {1,...,\numbersupplementpagesA}
    {
        \clearpage
        \includepdf[pages={\x}]{\supplementfilenameA}
    }
\fi

\ifarXiv
    \foreach \x in {1,...,\numbersupplementpagesB}
    {
        \clearpage
        \includepdf[pages={\x}]{\supplementfilenameB}
    }
\fi

\end{document}